\documentclass[12pt]{article}
\usepackage[latin9]{inputenc}

\usepackage{wrapfig}
\usepackage{units}
\usepackage{amsbsy}
\usepackage{amstext}
\usepackage{amsmath}
\usepackage{graphicx}

\usepackage{pdfpages}

\newcommand{\beginsupplement}{%
        \setcounter{table}{0}
        \renewcommand{\thetable}{S\arabic{table}}%
        \setcounter{figure}{0}
        \renewcommand{\thefigure}{S\arabic{figure}}%
     }

 \usepackage{times}

\newenvironment{sciabstract}{%
\begin{quote} \bf}{\end{quote}}

\begin{document}

\title{Loop Currents in Two-leg Ladder Cuprates}



\author{Dalila Bounoua$^{1\ast}$, Lucile Mangin-Thro$^{2}$, Jaehong Jeong$^{1}$, \\
Romuald Saint-Martin$^{3}$, Loreynne Pinsard-Gaudart$^{3}$, Yvan Sidis$^{1}$,\\
Philippe Bourges$^{1\ast}$\\
\\
\normalsize{$^{1}$ Laboratoire L\'eon Brillouin, IRAMIS/LLB, UMR12, CEA-CNRS, CEA-Saclay,}\\ 
\normalsize{Gif sur Yvette 91191, France}\\
\normalsize{$^{2}$ Institut Laue-Langevin, 71 avenue des martyrs, Grenoble 38000, France}\\
\normalsize{$^{3}$ Equipe Synth\`ese Propri\'et\'es et Mod\'elisation des Mat\'eriaux, Institut de Chimie Mol\'eculaire}\\
\normalsize{et des Mat\'eriaux d'Orsay, Centre National de la Recherche Scientifique UMR 8182,}\\
\normalsize{Universit\'e Paris-Sud, Universit\'e Paris-Saclay, 91405 Orsay, France}\\
\\
\normalsize{$^\ast$To whom correspondence should be addressed, e-mail: dalila.bounoua@cea.fr, philippe.bourges@cea.fr}
}

\date{}


 \baselineskip24pt



\maketitle

\begin{sciabstract}

 New phases with broken discrete Ising
symmetries are uncovered in quantum materials with strong electronic
correlations. The two-leg ladder
cuprate \textbf{$Sr_{14-x}Ca_{x}Cu_{24}O_{41}$} hosts a very rich
phase diagram where, upon hole doping, the system exhibits a
spin liquid state ending to an intriguing ordered magnetic state at
larger $Ca$ content. Using polarized neutron diffraction,  we report here the existence of
 short range magnetism in this material for two $Ca$
contents,  whose origin cannot be ascribed to Cu spins. This magnetism develops exclusively within the two-leg ladders
with a diffraction pattern at forbidden Bragg scattering which is
the hallmark of loop current-like magnetism breaking both time-reversal
and parity symmetries. Our discovery shows local discrete symmetry
breaking in a one dimensional spin liquid system as theoretically predicted.
It further suggests that a loop current-like phase could trigger the long range magnetic order reported at larger
doping in two-leg ladder cuprates. 

\end{sciabstract}


\section*{Introduction}

In the recent years, the study of doped Mott insulators, such as superconducting
(SC) cuprates \cite{fauque2006magnetic,li2008unusual,baledent2010two,bourges2011novel,mangin2015intra}
or iridates \cite{jeong2017time}, raised the question of the existence
of other kinds of magnetism. Beyond conventional spin magnetism, a
new form of magnetism may originate from magneto-electric loop currents
(LCs) \cite{simon2002detection,varma2006theory,agterberg2015emergent,chatterjee2017,chatterjee2017intertwining,scheurer2018orbital,sarkar2019incipient}
or Dirac multipoles \cite{lovesey2015neutron,lovesey2017neutron,spaldin2008toroidal,fechner2016quasistatic}.
While most of these states are usually discussed for hole doped quasi-2D
transition metal oxides such as cuprates and iridates, the existence
of LCs was also addressed in quasi-1D spin ladder cuprates \cite{chudzinski2008orbital,chudzinski2010spin,nishimoto2009current}.
\textbf{$Sr_{14-x}Ca_{x}Cu_{24}O_{41}$} is a prototype two-leg ladder
system whose hole doping can be tuned by $Ca$ for $Sr$ substitution.
This study is motivated by the recent theoretical proposal that spin liquids and topological order could be dressed with ancillary phases, such as LCs,
 that highlight their intrinsic nature \cite{chatterjee2017,chatterjee2017intertwining,scheurer2018orbital}. It represents a promising candidate for LCs hunting in the context
of low-dimensional spin liquids \cite{chudzinski2008orbital,chudzinski2010spin,nishimoto2009current}.

In cuprates, the LCs are expected to develop in the $CuO_{2}$ plaquettes,
the building blocks of the materials. In a 3-band Hubbard model, they
originate from the frustration of the electronic hopping and interaction
parameters and generate staggered orbital moments within the $CuO_{2}$
plaquettes. Once ordered, they are expected to preserve the lattice
translational invariance (q=0 magnetism) and break time reversal symmetry.
There may exist different LCs patterns with a single $CuO_{2}$ plaquette,
which can further break other $Z_{2}$ symmetries, such as parity
and rotation. In the so-called pseudo-gap phase of SC cuprates, there
are experimental evidences of a breaking of time-reversal, parity and rotation
symmetries, provided by polarized neutron diffraction (PND) \cite{fauque2006magnetic,li2008unusual,baledent2010two,bourges2011novel,mangin2015intra,de2012evidence,mangin2017,tang2018orientation},
muon spin spectroscopy \cite{zhang2018discovery}, 
second harmonic generation \cite{zhao2017global}, torque \cite{sato2017thermodynamic}
and optical birefringence measurements \cite{lubashevsky2014optical}. Further, second harmonic generation \cite{zhao2016}
and PND \cite{jeong2017time} observations in iridates provide evidence
for the universality of the LCs phase in correlated electron systems such as layered oxides.

In most SC cuprates, the LC state fully develops in the CuO$_2$ planes, yielding a 3D long range order. In lightly doped ${\rm La_{2-x}Sr_{x}CuO_{4}}$, known to host a spontaneous 
charge segragation in the form of quasi-1D charge stripes, the LC order is frustrated. As a result, the LC magnetism remains 
quasi-2D and  at very short range  \cite{baledent2010two}, suggesting a possible confinement of LCs within bond centrered stripes, taking the form of two-leg ladders.
However, the evidence for such a kind of LC-like magnetism in low-dimension remains
untackled up to date and can be investigated in the model quasi-1D
system $(Sr,Ca)_{14}Cu_{24}O_{41}$.

$Sr_{14-x}Ca_{x}Cu_{24}O_{41}$, hereafter $SCCO$-$x$, crystallizes
with an aperiodic nuclear structure consisting of an alternating stack
of 1D $CuO_{2}$ chains and quasi-1D $Cu_{2}O_{3}$ two-leg ladder
layers. It realizes an intrinsically hole-doped compound with an effective
charge of $2.25^{+}$ per Cu ion where, in the pure compound ($x=0$), 
holes are located within the chains subsystem. Substitution with $Ca^{2+}$
on the $Sr^{2+}$ site results in a charge transfer of the holes from
the chains to the ladders \cite{gotoh2003structural}, due to chemical
pressure. Owing to strong electronic correlations, hole doping strikingly
changes the electronic properties of $SCCO$-$x$, and the corresponding
phase diagram includes insulating gapped spin liquid phases within
the ladders ($\Delta_{gap}\sim32$meV), short-range dimer antiferromagnetic
orders within the chains, charge density wave in both chains and ladders,
magnetic long range order (LRO) - assumed to be antiferromagnetic
(AFM) - at large Ca-content, pressure induced superconductivity for
$\text{\ensuremath{x}\ensuremath{\ensuremath{\geq}}10}$, with a predicted
$d$-wave character in one-band Hubbard model \cite{Senechal2015},
and even pseudogap-like behavior for $x\geq9$ \cite{vuletic2006spin,uehara1996superconductivity}.
This work addresses the issue of the existence of LCs-like magnetism
in the archetypal hole-doped spin-ladders compounds, $SCCO$-$5$
and $SCCO$-$8$ with $\sim13$ and $\sim17\%$ hole doping per $Cu$
ion respectively, according to \cite{osafune1997optical}.

Using polarized neutron diffraction, we here report a short range magnetism in $SCCO$-$x$ below $\sim$50K and 80 K for two $Ca$ contents. 
This magnetism is associated exclusively with the two-leg ladder layers as the diffracted magnetic intensity occurs at forbidden Bragg positions of the ladder sub-system. Its origin cannot be related to Cu spins but rather may lie in the magnetoelectric loop currents, as the ones previously reported in superconducting 2D cuprates \cite{fauque2006magnetic,tang2018orientation}. The calculated magnetic structure factors, that satisfactorily reproduce our data,  correspond to two distinct loop current patterns proposed theoretically  \cite{simon2002detection,varma2006theory,chatterjee2017intertwining,scheurer2018orbital}. 
We further suggest that the long range magnetic order reported below $\simeq$ 5K at larger
doping in two-leg ladder cuprates could be induced by the observed short range magnetism. 

 \section*{Results}

\subsection*{Atomic structure}

\textbf{$Sr_{14-x}Ca_{x}Cu_{24}O_{41}$} exhibits an aperiodic atomic structure  (shown in Fig.\ref{Fig1}.a-b) 
with two, chains and  ladders, incommensurate sub-lattices \cite{deng2011structural}, which are described
 by the orthorhombic space groups \emph{Amma} and \emph{Fmmm}, respectively. 
The crystal lattice is incommensurate along the c-axis, with an incommensurability parameter 
between the chains and the ladders,
$\frac{1}{\gamma}=\frac{c_{Ladders}}{c_{chains}}\approx 1.43$. Upon Ca-doping(x$>$3),
the chains sub-space group changes from $Amma$ to $Fmmm$ such that
the whole structure is described  using 4D crystallography in Ref. \cite{deng2011structural}
as belonging to $Xmmm(00g)ss0$ superspace group, where X stands for
nonstandard centering $(0,0,0,0),(0,\frac{1}{2}$, $\frac{1}{2},\frac{1}{2})$, $(\frac{1}{2},\frac{1}{2},0,0)$ 
and $(\frac{1}{2},0,\frac{1}{2},\frac{1}{2})$.
In principle, Bragg peaks need then be indexed in the 4D superspace indexes
as $(H,K,L_{ladders,}L_{chains})$. However, as the reported magnetism
is basically related to the ladders sub-lattice, we refer through
this manuscript to the Bragg positions as (H,K,L), in units of: $\frac{2\pi}{a}=0.55\text{\AA}^{-1}$,
$\frac{2\pi}{b}=0.48\text{\AA}^{-1}$ and $\frac{2\pi}{c_{Ladders}}=1.61\text{\AA}^{-1}$,
with $a=11.4\text{\AA}$, $b=12.9\text{\AA}$ and $c=3.91\text{\AA}$, where
$L_{chains}=0$ (except where it is explicitly indicated otherwise).

Two large  $SCCO$-$x$ single crystals have been grown (see Methods) to perform the PND experiments, which are described in more details in
 the Supplementary Note \textbf{1}. During these experiments, 
 $SCCO$-$x$ samples were aligned with the ladder sub-lattice parameters
within the $(100)/(001)$ scattering plane so that wavevectors \textbf{Q
}of the form $(H,0,L)$ were accessible. Wavevectors are given in
reduced lattice units $(2\pi/a,2\pi/b,2\pi/c)$ where $a$, $b$,
$c$ stand for the lattice parameter of the ladder subsystem ($c$
along the legs, $a$, along the rungs and $b$ perpendicular to the
ladder planes $(ac)$, see Fig.\ref{Fig1}.a-b).

\subsection*{Short range magnetism in $SCCO$-$8$}

\begin{figure}
\begin{centering}
\includegraphics[width=10cm]{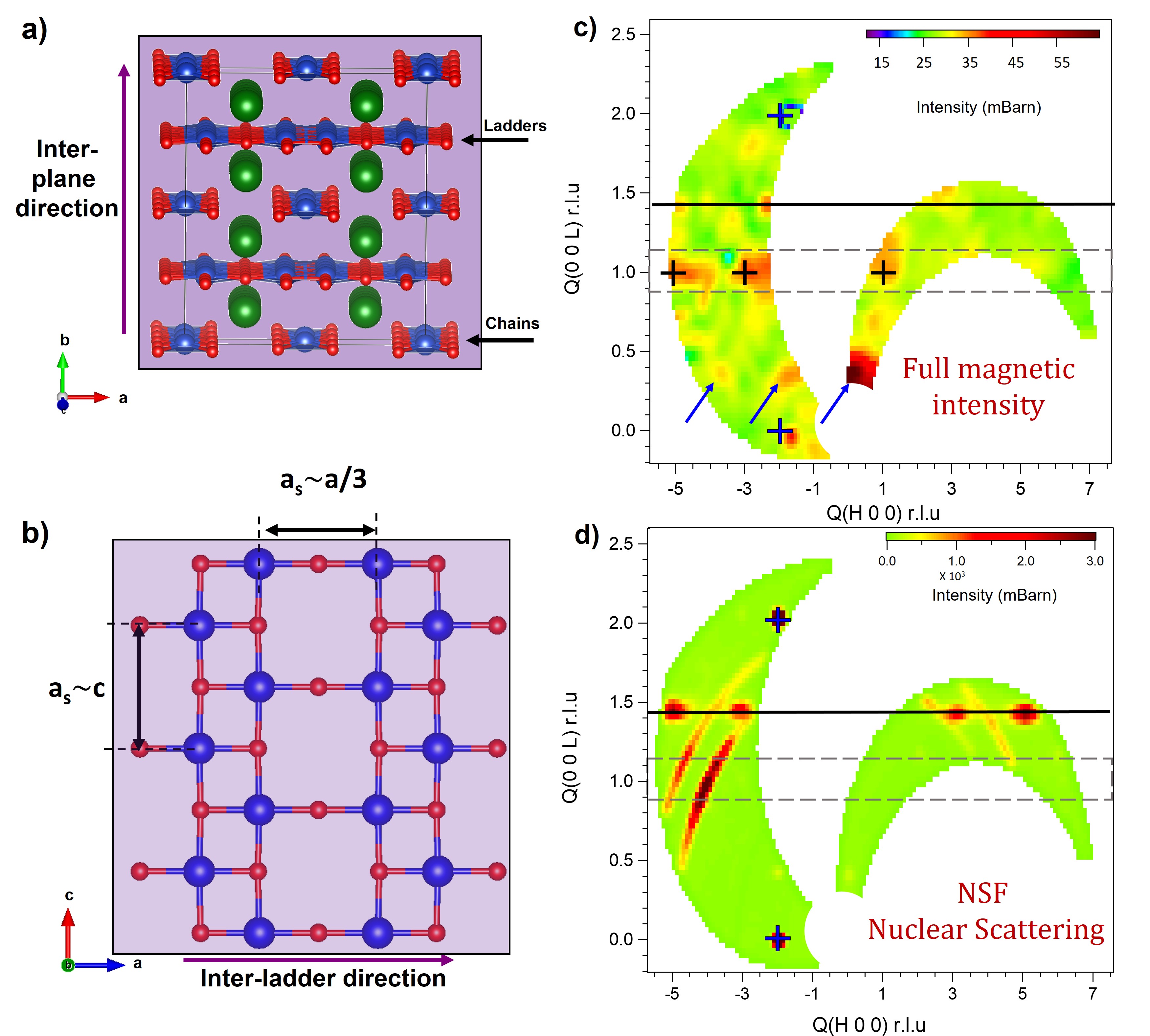} 
\par\end{centering}
\clearpage
\caption{\label{Fig1} \textbf{Crystal structure and magnetic map in $Sr_{6}Ca_{8}Cu_{24}O_{41}$,
sample}: \textbf{(a) }Crystal structure of $Sr_{14}Cu_{24}O_{41}$
showing the alternating stack of $CuO_{2}$ chains (Cu in blue and
O in red) and $Cu_{2}O_{3}$ ladder planes, separated by $Sr$ ions
(in green) along the b-axis. \textbf{(b) }{[}a,c{]} plane projection
of the ladders subsystem including one two-leg ladder. The ladders are formed by edge sharing $CuO_{2}$\textbf{
}squares.  \textbf{(c-d)} Mapping in momentum space of the full magnetic scattering
deduced from XYZ-PA \textbf{(c)}, and nuclear intensity measured in
the non spin-flip (NSF) channel \textbf{(d) }(measured on $D7$, $T=5K$). The maps\textbf{
}are given in reduced lattice unit (r.l.u.) of the ladders subsystem and the intensities in
mbarn. The area bounded by dashed lines indicates the ladder scattering
ridge along (H,0,1). The solid lines are associated with the chains
nuclear response and blue crosses at integer H and L values correspond
to the nuclear Bragg scattering associated with the ladders : \textbf{(c)
}The ladders and satellite reflections magnetic spots are located
by crosses and blue arrows, respectively.\textbf{ }The magnetic satellite
reflections are of the form $(H,0,-1,1)$ using 4D superspace notations. In \textbf{(d)}, the halos correspond
to aluminum powder scattering from the sample holder. Sharper red spots in \textbf{(c)} 
located at positions of strong nuclear Bragg peaks in the NSF map  \textbf{(d)}
are not of magnetic origin but corresponds to polarization 
leakages from the NSF channel.}
\end{figure}

 Fig. \ref{Fig1}.c shows a \textbf{Q}-map of the full magnetic scattering at $T=5K$ in $SCCO$-$8$,
measured on the D7 diffractometer extracted from longitudinal XYZ-PA
(see Methods). A magnetic signal is systematically observed
along (H,0,1) for odd integer H values and where L=1 corresponds to
the ladders sublattice (crosses on Fig.\ref{Fig1}.c). These \textbf{Q}-positions
correspond to nuclear extinctions according to the atomic space group
symmetry selection rules as shown by the absence of nuclear
scattering at the same \textbf{Q}-positions in the NSF channel Fig.\ref{Fig1}.d.
However, the magnetic scattering appears significantly broader than the instrumental
resolution, the hallmark of only short range magnetism.

\begin{figure}
\begin{centering}
 \includegraphics[width=14cm]{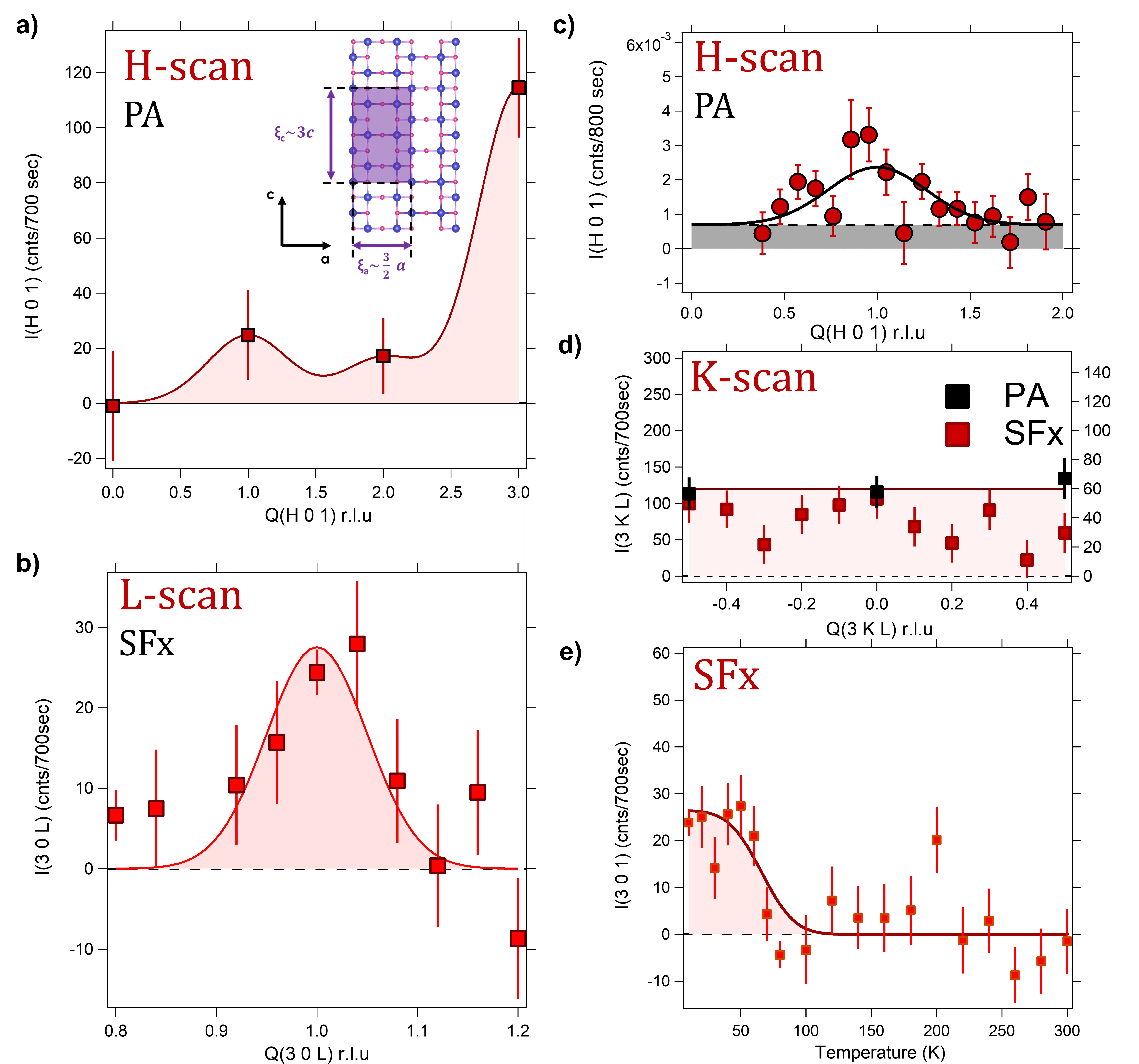} 
\par\end{centering}
\centering{}\caption{\label{Fig2} \textbf{Magnetic scans in  $Sr_{6}Ca_{8}Cu_{24}O_{41}$}: 
  \textbf{(a)}   Magnetic intensity of (H,0,1) points for integer H, {\it i.e.} along 
 the inter-ladders direction (rungs) as extracted from XYZ polarisation analysis (XYZ-PA). 
The inset shows real-space magnetic correlation lengths within the ladders planes.
 \textbf{(b)}  Background substracted L-scan across
(3,0,1) in the spin-flip ($SF_{X}$) channel. The magnetic intensity appears as
a Gaussian signal centered at (3,0,1) (Raw data given in Supplementary
Note \textbf{2}). \textbf{(c)}  H-scan across (1,0,1) direction,
along the inter-ladders direction (rungs) extracted from  (XYZ-PA). \textbf{(d)}
K-scan across (3,0,1) showing the magnetic intensity along the inter-plane
direction as deduced from  XYZ-PA (black) and $SF_{X}$ measurements
after subtraction of a background intensity taken at (3,K,0.8) (red).
\textbf{(e)} Temperature dependence of the magnetic intensity at (3,0,1),
measured in $SF_{X}$ and obtained after background subtraction (Raw
data given in Supplementary Note \textbf{3}). Data in \textbf{(c)}
were measured on $D7$ at $5K$ and the others on $4F1$ at $10K$.
Lines are guide to the eye.  Error bars represent one standard deviation.  }
\end{figure}

All of these results, obtained on the D7 diffractometer, are confirmed
on the TAS-4F1 $(T=10K)$. Fig.\ref{Fig2}.a  show the measured magnetic
scattering, as extracted from XYZ-PA across the inter-ladder direction
(H,0,1) in agreement 
with the results from D7. The \textbf{Q}-dependence of the magnetic
intensity exhibits a peculiar structure factor with the absence of
scattering for $H=0$ and an enhanced intensity at $H=3$, along the
ladders scattering ridge. To better characterize this  short range magnetism (SRM), we performed
scans across selected positions of the (H,0,1) rod. A L-scan across
(3,0,1) position ($SF_{X}$, 4F1, $T=10K$) (raw data are shown in Supplementary Note \textbf{2}),
 and a H-scan across (1,0,1) as extracted from
the D7 XYZ-PA map ($T=5K)$, are reported in Fig.\ref{Fig2}.b and  Fig.\ref{Fig2}.c,
respectively. The scans show peaked signals with momentum widths (FWHM), $\Delta_{H}=0.5\:r.l.u$ and $\Delta_{L}=0.12\,r.l.u$,
which are much broader than the instrumental resolution. All along the manuscript, the correlation lengths along both directions have been deduced after deconvolution from the instrument resolution,  as $\xi_{a}=\frac{a}{\pi \Delta^{dec}_{H}}$ and $\xi_{c}=\frac{c}{\pi \Delta^{dec}_{L}}$.
The corresponding correlation lengths along the inter-ladder direction
is $\xi_{a} \sim7\pm2.5\text{\AA}$  equivalent to
$a/2$. $\xi_{a}$ corresponds to the size of one ladder
rung (2 $\times$ Cu-O bond lengths) plus two inter-ladders spacings
(2 $\times$ Cu-O bond length) as shown in the inset of Fig.\ref{Fig2}.a. $\xi_{a}$
along the ladder legs is found to be $\xi_{c}\sim11\text{\ensuremath{\pm3}\AA}$,
or correspondingly $\sim3c$. The {[}a,c{]} in-plane correlation lengths
are very short range and indicative of the formation of magnetic
clusters\emph{ }within the ladders.

Next, we performed a survey of the K-dependence of the magnetic scattering,
along the inter-planes direction (Fig.\ref{Fig2}.d). We collected
a $SF_{X}$ scan for a trajectory of the form (3,K,L) with $L=0.8$
and $1$. The scan at $L=0.8$ stands for a nonmagnetic background
according to Fig.\ref{Fig2}.b. Subtracting $L=0.8$ from $L=1$ data
unveils a roughly constant level of magnetic intensity over the measured
K-range, in agreement with XYZ-PA data (Fig.\ref{Fig2}.e). This indicates
vanishing inter-plane correlations, emphasizing a 2D confinement
of the measured magnetism within the ladder planes. The temperature
dependence of the magnetic signal (Fig.\ref{Fig2}.e), measured at
(3,0,1), in the $SF_{X}$ channel (4F1)  (raw data are shown in Supplementary Note \textbf{3}), 
shows that the magnetic correlations set-in
below $T_{mag}\sim80K$.

\begin{figure}
\begin{centering}
 \includegraphics[width=14cm]{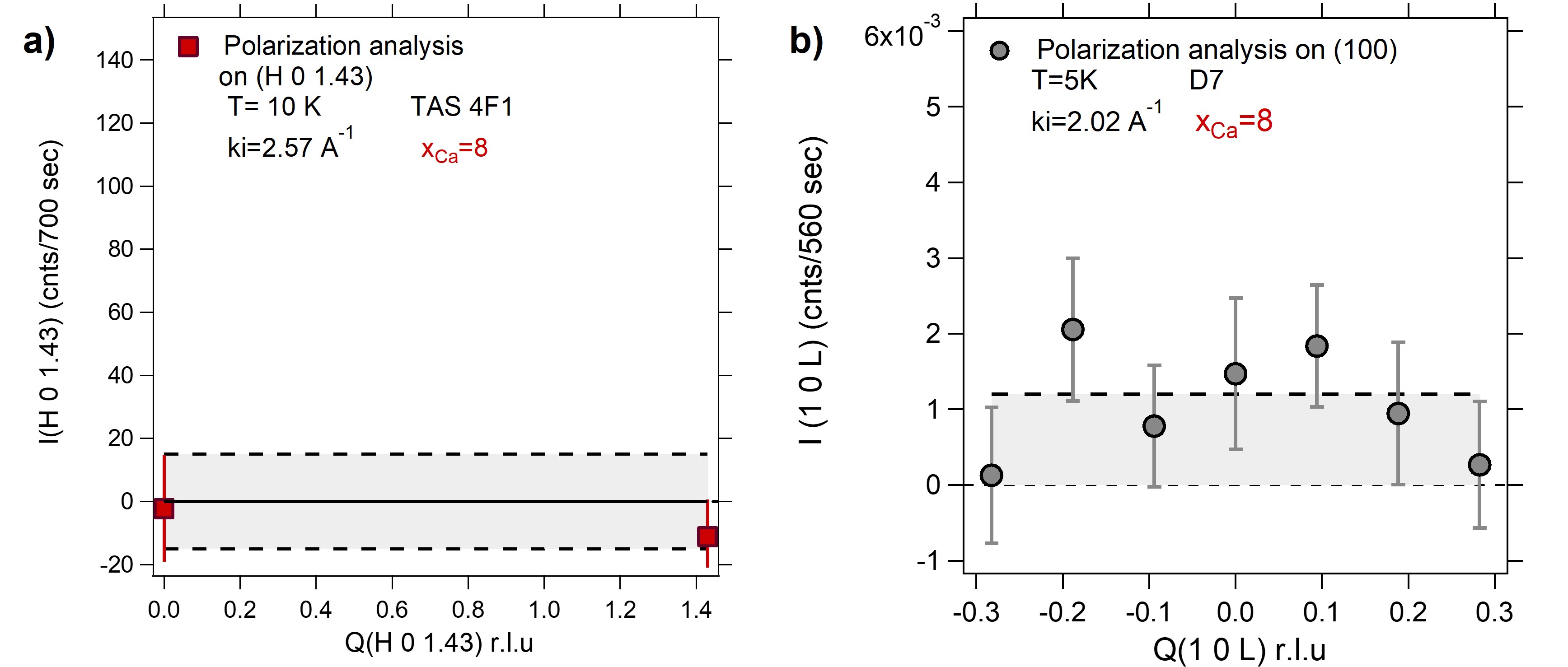}
\par\end{centering}
\caption{ \textbf{Absence of magnetic signal on chains in  $Sr_{6}Ca_{8}Cu_{24}O_{41}$}: Full magnetic intensity deduced from XYZ polarisation analysis on (a)  (4F1, T=10K) within the chain subsystem along (H,0,0,1) in superspace reduced lattice unit (r.l.u.) or (H,0,1.43) in ladders r.l.u. (b) (D7, T=5K) along (1,0,L). Error bars represent one standard deviation. 
}
\label{fig:Chains} 
\end{figure}

A new magnetic signal is clearly observed at $(H,0,L)$ with integer values of $H$ and $L$, corresponding to the ladder subsystem.
This raises the question of its possible existence (or fingerprint)
also in the chains subsystem.  In order to answer that question, one needs to look 
at Bragg positions  $(H,0,0,1)$ using superspace notations,  corresponding
to $(H,0,1.43)$ in the ladder subsystem units. In  Fig. \ref{Fig1}.c, no magnetic intensity is sizeable at these positions. 
Furthermore, Fig~\ref{fig:Chains}.a, reporting the full scattered magnetic intensity as deduced from XYZ-PA
(4F1) at two Bragg positions, shows  no magnetism. That reveals
the absence of any chains magnetic response at commensurate chains
Q-positions. A scan across the $(1,0,L)$ direction in $SCCO-8$ was
also performed in order to crosscheck the absence of signal at positions
of the form $(H,0,0)$ (common to both ladders and chains), as shown in Fig.~\ref{fig:Chains}.b. All
scans reported in Fig.~\ref{fig:Chains}, emphasize that the SRM detected within
the  Cu$_2$O$_3$ ladders is not transferred to the underlying CuO$_2$ chain subsystem.

However, a correlated magnetic signal appears on top
of satellite reflections mixing both ladders and chains subsystems, along
the $(H,0,0.43)$ line, with H-even (blue arrows Fig.\ref{Fig1}.c),
in r.l.u of the ladders, corresponding to the (H,0,-1,1) line in the
superspace notations. That unexpected result does not necessarily mean that the chain subsystem carries magnetic moments
 which would be inconsistent with the absence of a magnetic signal on the chains Bragg positions. Instead,  that observation suggests
that the magnetic moments of the ladder could actually be magnetically coupled via the chain susbsytem, leading to a non-zero magnetic structure factor at the satellite.  
That interpretation is inline with the inter-ladder correlations of the magnetic signal. 

\subsection*{Lost inter-ladder correlations at lower Ca-substitution: $SCCO$-$5$}

Fig.\ref{Fig3}.a shows a L-scan in the SF channel across (0,0,1)
along the ladder-legs direction ($SF_{X}$, 4F1, $T=5K$), 
where no magnetic signal*
is observed for $SCCO$-$8$. The scan shows a clear magnetic signal
centered at (0,0,1), where nuclear scattering is forbidden. The FWHM
of the measured signal gives (after deconvolution from the instrumental resolution) a correlation length of $\xi_{c}\sim20\pm6\text{\AA}$
along the ladders or equivalently $\sim6c$, which is enhanced as
compared to the $SCCO$-$8$ compound. We further performed an XYZ-PA
on 4F1 along the inter-ladder direction (H,0,1). The XYZ-PA reported
in Fig.\ref{Fig3}.b reveals a diffuse magnetic scattering along $H$
indicating a vanishing $\xi_{a}$ with only a minimum of the magnetic
intensity for $H=1.5$. These results highlight an even shorter-range
magnetism, confined within a single ladder, and the loss of inter-ladder
magnetic correlations when lowering the $Ca$-content. 
Two features, found in $SCCO$-$8$, have been similarly observed.
First, the XYZ-PA along (3,K,1) reveals no maximum intensity at $K=0$. In light of the results for x=8 sample (Fig.\ref{Fig2}.d), 
this is consistent with the absence of correlations along the b-axis (Fig.\ref{Fig3}.c), the
inter-planes direction.  Second, the XYZ-PA within the chains subsystem confirms
the absence of chains magnetism (Supplementary Note \textbf{4}).  However, 
in contrast to  $SCCO$-$8$, no magnetic intensity is observed at the satellite 
position  corresponding to the (H,0,-1,1) line in the superspace notations, confirming the assumption
 that the CuO chain could bridge inter-ladder coupling in  $SCCO$-$8$.

 Fig.\ref{Fig3}.d shows the temperature
dependencies of the magnetic intensity at (3,0,1) and (1,0,1), respectively
(4F1). The signal at (3,0,1) was measured in the $SF_{X}$ channel
and corrected from a background intensity measured at (3,0,0.8). The
magnetic signal at (1,0,1) was tracked as a function of the temperature
using unpolarized neutrons. Both datasets give an onset temperature
$T_{mag}\sim50K$.

Put together, all these experimental observations allow one to get
a rather accurate description of the observed magnetic patterns, especially
thanks to the large set of collected magnetic instensites at various
Q points: i) The magnetic signal is short-range, 2D and exclusively
carried by the ladder subsystems with weak inter-ladders correlations.
ii) The magnetic scattering appears on wavevectors of the form (H,0,L)
with integer and odd H and L values, which are forbidden for the atomic
structure due to additional symmetries of the 3D crystal structure
\cite{deng2011structural}. That indicates that the translational
invariance of the ladders sub-lattice is preserved with the same magnetic
unit cell as the atomic one (q=0 magnetism), as reported for the superconducting
cuprates and iridates \cite{fauque2006magnetic,baledent2010two,bourges2011novel,mangin2015intra,jeong2017time},
which is usually interpreted in terms of LCs. These first two points
concern both Ca contents. In contrast to the $SCCO$-$8$ compound
where the magnetic intensity exhibits a pronounced maximum at (3,0,1),
the SRM remains confined to a single two-leg ladder for $SCCO$-$5$, as reported for the $(La,Sr)_{2}CuO_{4}$
cuprate \cite{baledent2010two}, with only a minimum intensity at H=1.5.
Similarly, the observation of magnetic intensity at the satellite position  (H,0,-1,1) differs noticeably between both Ca-concentrations
inline with the loss of the inter-ladder correlations at low Ca-substitution. 


\subsection*{Amplitude of the SRM}

 The scattering intensity can be converted
in absolute units (barn), after a calibration using a reference vanadium
sample (Supplementary Note \textbf{5}). This leads to a full
magnetic scattered intensity of $I_{mag}\sim28\pm4\,mbarn$ and $I_{mag}\sim36\pm15\,mbarn$
for $SCCO$-$8$, on 4F1 and D7 respectively, at $(3,0,1)$, where
the structure factor is maximum. Correspondingly, the full magnetic
intensity was found to be $I_{mag}\sim7\pm2\,mbarn$ for $SCCO$-$5$
at the same wavevector. These amplitudes correspond to the scattered
magnetic intensity of one $(Sr,Ca)_{14}Cu_{24}O_{41}$ formula unit
(f.u), namely, three $CuO_{2}$ square unit cells with $4\,Cu$/f.u
(insets of Fig.\ref{Fig4}.a). Once normalized to a single $Cu$ site, these
amplitudes remain larger than those reported in superconducting cuprates
($I_{mag}\sim$1-2 mbarn per $Cu$) \cite{fauque2006magnetic,baledent2010two,bourges2011novel}.
To date, we report the largest intensity of the SRM respecting invariance symmetry
(q=0) in cuprates.

\subsection*{Orientation of the magnetic moments} 

The magnitude and orientation
of the measurable SRM magnetic moment $m$ is defined as $m^{2}=m_{ac}^{2}+m_{b}^{2}$,
where $m_{ac}$ and $m_{b}$ denote the ladders in-plane and out-of-plane
magnetic moment, respectively (Supplementary Note \textbf{6})
and $m_{ac}^{2}=m_{a}^{2}+m_{c}^{2}$. Both components,  $m_{ac}$ and $m_{b}$, can be derived
from a full XYZ-PA but not the values  $m_{a}$ and $m_{c}$ which are interdependent. Supposing that $m_{a}=m_{c}$, we reproducibly
estimate the ratio $(\frac{m_{b}}{m_{ac}})^{2}\sim1$ for both compounds.
Consistently, 4F1 and D7 data show that 50\% of the magnetic moment
lies out of the ladder planes, with a tilt of the out-of plane magnetic
moment to an angle $\Theta=Atan(\nicefrac{m_{b}}{m_{ac}})\sim$ 55
$^{\circ}$. This is in agreement with previous estimates in SC cuprates
where the magnetic moment associated with the LCs magnetism exhibits
a similar tilt \cite{fauque2006magnetic,mangin2015intra,tang2018orientation}.

\section*{Discussion}

{\bf Phase diagram}:  Our experiments highlight the systematic onset of SRM within
the Cu-O planes of lightly hole-doped spin ladders, with growing correlations
upon increasing the hole content (Fig.\ref{Fig4}.a). We note as well the absence of a structural distorsion associated with a charge 
density wave instability in our samples (see Supplementary Note \textbf{7}). According to
magnetic susceptibility and specific heat data, no phase transition
occurs in this region of the phase diagram \cite{vuletic2006spin}. 
However at larger Ca doping ($x\geq9$), an AFM LRO phase is reported,
but only below 4.2K (Fig.\ref{Fig4}.a) whose antiferromagnetic nature
has been basically deduced through the cusp in the temperature dependence
of the macroscopic susceptibility \cite{isobe1999antiferromagnetic,nagata1999antiferromagnetic,nagata2001antiferromagnetic,deng2013coexistence}.
Interestingly, the diffraction patterns of the SRM and the AFM order are located at the same wave vector in momentum space, suggesting a related
origin. As the SRM occurs at higher temperature, it is tempting to
propose that the reported SRM could act as a preemptive state of the
AFM LRO as the Ca-doping evolution of the correlations suggests.

Meanwhile, the instrinsic nature of the AFM-LRO remains under debate as the reported locations
of the magnetic Bragg peaks do not correspond to any simple model
of antiferromagnetically interacting Cu spins within the ladder legs
or the chains. Neutron diffraction data on single crystals indicate
that the LRO involves magnetic moments both in the ladders and the
chains subsystems \cite{nagata1999antiferromagnetic,nagata2001antiferromagnetic,deng2013coexistence}
as it gives magnetic scattering at integer $H$ and $L$ of both ladders
and chains sublattices. Therefore, complex Cu spin structures, which
typically require to consider large super cells with a considerable number
of independent spins, has to be invoked to describe the magnetic diffraction
patterns. As recognized by Nagata {\it et al}\cite{nagata1999antiferromagnetic,nagata2001antiferromagnetic},
the model although reproducing the experimental data gives rise to
an unlikely situation where the magnetic interaction between nearest
neighbor Cu spins are either ferromagnetic or antiferromagnetic and
these interactions are mixed with some periodicity. Deng {\it et al} \cite{deng2013coexistence} are reporting
two different highly non trivial magnetic structures 
with 96 independent spins in a large super cell (corresponding to
$24\:SCCO$ unit cells) to accommodate the measured structure factors
of 36 magnetic Bragg peaks. Obviously, such a model fails  to produce
the $SRM$ of $SCCO$-$5$ and $8$ as the reported correlation lengths
are much shorter than the super-cell size necessary to describe the complex spin structure. 
Additionally, the orientation of the magnetic moments in the AFM LRO phase, assumed to be related to spin
moments, is  found to be strictly in-plane (in the  [a,c] plane) according to neutron diffraction data \cite{nagata1999antiferromagnetic,nagata2001antiferromagnetic}.
That contrasts with the observed SRM and suggests that both phases are distinct although related. 

{\bf Modelisation}: We next propose  various magnetic models which could account for the observed magnetic intensities as shown in Figs.\ref{Fig4}.b-c.  One then needs to calculate the magnetic structure factor expected  for given magnetic patterns. This is discussed in details in the Supplementary Note \textbf{8} for various magnetic configurations; we here recall the main conclusions.  We consider various decorations of the ladder unit cell as shown in Fig.~\ref{patterns}. Some models have been proposed theoretically, others are simplified magnetic decorations of the unit cell.

 First, one considers a model of periodic antiferromagnetic  Cu spins on a square lattice  (Fig.~\ref{patterns}.a), that can decorate an  isolated ladder.  In principle, this model can be directly applied 
to $SCCO$-$5$ sample where one observes a magnetism confined to a single ladder.  However, it completely fails to reproduce our experimental
data as it breaks the translation symmetry of the lattice and would lead to SRM with scattering only at half integer
$H$ and $L$ values at odds with our observation.  For the same reason, any model built of such an AFM coupled ladder does not 
describe the data whatever the coupling between ladders.  Neither does a model of magnetic  (spin or orbital) moments on oxygen
sites as considered in \cite{fauque2006magnetic,sun2009topological,fischer2011mean,moskvin2012pseudogap}
(Fig.~\ref{patterns}.b). That model does respect the translation of the lattice but leads to an extinction of the structure factor at (3,0,1) where the observed intensity is maximum. 
Therefore, one is left to find out alternative models to give
an explanation for the SRM and a possible link with the AFM LRO.

{\bf Loop Currents (LCs) modeling}: In a marked contrast with highly complex magnetic arrangements of
$Cu$ spins used to describe the LRO AFM state, we propose a comprehensive interpretation of our
PND measurements in the framework of LCs in two-leg ladders. Following
theoretical proposals \cite{simon2002detection,varma2006theory,chatterjee2017intertwining,scheurer2018orbital},
we calculate two magnetic structure factors corresponding to two distinct
LCs patterns that satisfactorily reproduce our data. These two patterns
are based on a set of two counter-propagating LCs per $Cu$ site.
At variance, other patterns with a set of four LCs (Fig.~\ref{patterns}.c usually referred
to as $CC-\theta_{I}$ phase) \cite{chudzinski2008orbital,chudzinski2010spin,nishimoto2009current}
give rise to different magnetic scattering selection rules that do
not satisfy the measured structure factor with extinctions at Bragg positions where the observed intensities are 
maximum. 


{\bf CC-$\Theta_{II}$ model}: The first model consists in a CC-$\Theta_{II}$ like pattern of LCs  (Fig.~\ref{patterns}.d)
\cite{simon2002detection,varma2006theory} within the ladders. One needs to decorate each ladder unit cell
(insets of Fig. \ref{Fig4}.a) of $\sim$ 3 Cu-O plaquettes (each plaquette
has an averaged cell parameter of $a_{s}=c\sim a/3$ as shown in Fig.\ref{Fig1}.b)
with the two opposite LCs around each Cu atom. Note that only the
out-of-plane magnetic component $m_{b}$ perpendicular to the LCs, $\equiv$ $m_{LC}$,  can be considered for modeling \cite{mangin2017}. 
Further, one considers equal contribution from the 4-fold degenerate domains given by a 90$^{\circ}$
rotation of LCs about the Cu-site \cite{varma2006theory,bourges2011novel}. The $Cu$ magnetic form factor was used
 to fit the experimental data shown in  Fig.\ref{Fig4}.b-c. Note that the data could be as well described by using an oxygen form factor
(see  Supplementary Note \textbf{9}).

First, the model of isolated ladders with the same pattern around each Cu atom (inset A of Fig.\ref{Fig4}.a) nicely reproduces the main features of the $SCCO$-$5$ data  along the (H,0,1) line (Fig.\ref{Fig4}.b) accounting for lost inter-ladders correlations. It explains also why the magnetic correlations along \textbf{a} (perpendicular to the ladders)  are confined within a single ladder. The model gives a $m_{LC}=0.05\pm0.01\mu_{B}$ estimate for the magnetic moment amplitude (see  Supplementary Note \textbf{10}). Next, we add inter-ladder correlations which are typically imposed by the structure geometry and currents continuity as shown 
in the inset B of Fig.\ref{Fig4}.a. Again, the same model reasonably reproduces the data of $SCCO$-$8$ shown by Fig.\ref{Fig4}.c, 
with a  comparable magnetic moment amplitude of $m_{LC}=0.05\pm0.01\mu_{B}$.  For both samples and although
the magnetic cross section is larger than in superconducting cuprates,
the LCs magnetic moment is of the same order of magnitude \cite{fauque2006magnetic,baledent2010two,bourges2011novel,mangin2015intra}
due to more complex magnetic structure factor with interferences related to the larger
magnetic unit cell and, as well, because only $m_{b}$ is here considered.


{\bf CC-$\Theta_{III}$ model}: Next, we test the model proposed in refs. \cite{chatterjee2017intertwining,scheurer2018orbital}
for the case of two-dimensional spin-liquids, here labeled $CC-\Theta_{III}$  constructed from patterns shown in Fig.~\ref{patterns}.e. 
As for CC-$\Theta_{II}$, there are two counter-propagating LCs, but the currents now flow only
between O-sites (insets of Fig.\ref{Fig4}.a). Typically, the LCs pattern is
rotated by 45$^{\circ}$ with respect to CC-$\Theta_{II}$. Similarly,
two orientational domains rotated by 90$^{\circ}$ about Cu atoms,
with a two-fold degeneracy each, could be considered for this phase.
However, only one orientation is found to give agreement with the experimental data along the (H,0,1) line of uncorrelated $SCCO$-$5$ (dashed line in Fig.\ref{Fig4}.b). 
It is  the \textit{Vertical}-$CC-\theta_{III}$ pattern,  represented in the inset A of
Fig.\ref{Fig4}.a.  The \textit{Horizontal}-$CC-\theta_{III}$ pattern cannot explain the measured 
data at H=0. In the case of correlated ladders $SCCO$-$8$, both patterns can describe the 
data of Fig.\ref{Fig4}.c.  The fits give
good agreement to the experimental data yielding $m_{LC}$ magnitudes
of $m_{LC}=0.04\pm0.01\mu_{B}$ and $m_{LC}=0.04\pm0.01\mu_{B}$ for
$SCCO-8$ and $SCCO-5$, respectively. 

  Interestingly, for  isolated ladders, 
only one orientational domain of $CC-\Theta_{III}$ can account for the experimental data
in contrast with  the $CC-\theta_{II}$ model which has no preferred domain orientation.
In terms of the observed magnetic structure factors, magnetic dipoles (as Cu spins or moments at the oxygen sites) fail to describe our data. This is a clear indication that higher multipoles, ordered at short range, are necessary. It can correspond to anapoles (such as the LCs phases) or magnetic quadrupoles as both are intimately
connected as they occur at the same level of the multipole expansion\cite{dimatteo2012}.  Both types of order parameters break both parity and time-reversal symmetries and would,  
in principle,  cause the measured magnetism. For instance, and from purely symmetrical considerations, it is likely that models built from magneto-electric quadrupoles, as the ones proposed in 2D cuprates \cite{lovesey2015neutron,lovesey2017neutron,fechner2016quasistatic}, could describe the measured magnetism if proper couplings between adjacent quadrupoles in the ladder unit cell are considered. However, at present, it is unclear what type of microscopic couplings between quadrupoles would correlate adjacent ladders. We here stressed that LCs phases offer more documented microscopic models where magnetic structure factors can be readily computed. 


Our report is inline with the observation of loop currents in 2D superconducting cuprates  \cite{fauque2006magnetic,li2008unusual,baledent2010two,bourges2011novel,mangin2015intra}. Particularly, it shows that LCs phases occur as well in 1D spin liquids systems as it has been theoretically anticipated  \cite{chatterjee2017intertwining,scheurer2018orbital,chudzinski2008orbital,chudzinski2010spin,nishimoto2009current}. Consistently, it is interesting to notice that  short-range orbital-like magnetic order has been as well reported in ${\rm La_{2-x}Sr_xCuO_4}$ once the doped charges are confined in two-leg ladders. \cite{baledent2010two}, bridging  the occurence of loop currents in various copper oxides. 
The quasi-1D structure of Cu$_2$O$_3$ ladder  enforces geometrical constraints that allow us to be more specific about the LCs models because the ladder unit-cell is anisotropic and it exhibits a different atomic structure. For instance, although there is an oxygen atom above Cu atoms on the ladders \cite{isobe1998} at about the same distance, typically $\sim$ 2.7 \AA,  as the apical oxygen
in hole doped  mono- or bi-layer cuprates where copper is located within a CuO$_6$ octahedron or CuO$_5$ pyramid, the atomic structure is aperiodic and then that oxygen does not primarily belong to the same atomic sub-system than the Cu atom on the ladder.  Therefore, one can consider that  LCs are established only in the planar CuO$_2$ plane, that is the first time LCs correlations are observed without apical oxygen. That point is particularly the case in the $SCCO-5$ sample of decorrelated ladders where no magnetic intensity occurs at the satellite position, (0,0,-1,1), mixing both sublattices. The occurrence of LCs in a system without specific apical oxygen has stringent consequences to explain the observed 
tilt of the q=0 magnetism \cite{tang2018orientation} (see Supplementary Note  \textbf{11}). 

Both, CC-$\Theta_{II}$ and CC-$\Theta_{III}$, LCs models that we propose capture the most salient observation of
the \textbf{Q}-dependence of the magnetic scattered intensity as reported
in Fig. \ref{Fig4}.b and c whereas other models based on magnetic moments on Cu or oxygen sites fail to account for the experimental results. 
Note that both LCs patterns can simply be generated from a single
LC orientation and considering the lattice symmetry at variance with
the proposed magnetic super-cell with 96 spins to describe LRO AFM
phase\cite{deng2013coexistence}. In our previous work on LSCO, the appearence of the LC-like magnetism 
also coincides with a net anomaly in the spin dynamics,  suggesting an interplay between AFM spin correlation 
and LC-like electronic instability\cite{baledent2010two}. A similar interplay could be present in two-leg ladders, 
so that LC-like phase triggers at low temperature  the AFM order at large Ca content.
This would be consistent with a picture of a fluctuating N\'eel state
(spin liquid state) carrying preemptive LCs orders, by analogy to
the LCs order parameter resulting from the intertwining between a
topological order and discrete broken symmetries in 2D spin liquids
\cite{chatterjee2017,chatterjee2017intertwining,scheurer2018orbital}.

\section*{Acknowledgments}

The authors would like to acknowledge  B. Fauqu\'e,  M. Greven, A. Goukassov, C.M. Varma and A. Wildes 
for fruitful discussions. We acknonwledge F. Maignen
and C. Meunier for their valuable technical assistance on 4F1. We
acknowledge supports from the project NirvAna (contract ANR-14-OHRI-0010)
of the French Agence Nationale de la Recherche (ANR). The open access fee was covered by FILL2030, a European Union
project within the European Commission's Horizon 2020 Research and
Innovation programme under grant agreement N$^{\circ}$731096.

\noindent \textbf{Author contributions.} Y.S. and  P.B. conceived and supervised the project; D.B., Y.S. and P.B. performed the experiments at LLB Saclay  with support from J.J.;  D.B. and L. M.-T. performed the experiments at ILL Grenoble; D.B. and P.B. analysed the neutron data; R.S.-M. synthesized the single crystal samples with support from L P.-G.; D.B., Y.S. and P.B. wrote the manuscript with further contributions from all authors. All authors contributed to this work, read the manuscript and agree to its contents. \\ \\
\noindent \textbf{Competing Interests.} 
The authors declare no competing interests.\\ 
\noindent \textbf{Data Availability}. Data collected on D7 are available at  https://doi.ill.fr/10.5291/ILL-DATA.5-53-279. The rest of the data that support the findings of this study is available from the corresponding authors upon request. \\

 \section*{Methods}

 \section*{Crystal growth} 

We report a PND study of two $SCCO$-$x$ single crystals. The single crystals of \textbf{$Sr_{14-x}Ca_{x}Cu_{24}O_{41}$}
with ($x=5$ and $8$) were grown by the traveling solvent floating
zone method \cite{revcolevschi1999} using a four mirrors image furnace at SP2M-ICMMO. The
crystals were grown from polycrystalline feed rods of the corresponding
compounds, obtained by solid state reaction of stoichiometric amounts
of $CuO$, $SrO$ and $CaO$ precursors. The growth was carried out
under an oxygen pressure of 5 and 8 bars for the x= 5 and 8 $Ca$-doped
$Sr_{14}Cu_{24}O_{41}$ respectively, in order to avoid the formation
of secondary phases and favor the constrained structure resulting
from Ca-doping. The growth was initiated using a solvent pellet containing
$30\%$ $(Sr,Ca)O$ and $70\%$ $CuO$ and carried out with a rate
of $1mm.h^{-1}$.The crystals weigh $2.1$g and $2.5$g for the $x=5$
and $8$ compositions, respectively.


\section*{Polarized neutron diffraction}

The PND experiments of $SCCO$-$x$ single crystals,  described in more details in
 the Supplementary Note \textbf{1}, were carried out on two instruments: the triple axis spectrometer (TAS)
4F1 (Orph\'ee reactor, Saclay) and the multidetector diffractometer
D7 (Institut Laue Langevin, Grenoble).  These instruments are equipped with distinct neutron
polarization set-ups and were operating with two distinct neutron
wavelengths, to guarantee the reproducibility of the measurements.
In a PND experiment, the quantization axis of the neutron spin polarization,
$\boldsymbol{P}$, is given within a (X,Y,Z) Cartesian referential.
For a fixed neutron spin polarization, one can selectively probe the
scattered intensity in the spin flip channel ($SF_{P}$) where the
neutron spin polarization is reversed after interaction with the sample,
and the non-spin-flip channel ($NSF_{P}$), where the neutron spin
is conserved. The amount of the magnetic scattering in the $SF_{P}$
channel varies as a function of the polarization $\boldsymbol{P}$.
The combination of measurements in the $SF$ channel for different
$\boldsymbol{P}$ is called the polarization analysis (PA). It allows
a full determination of the magnetic intensity $I_{mag}$ where the
non-magnetic background is removed. 

We used incident neutron wavevectors
of ${\bf k_{i}}=2.57 \text{\AA}^{-1}$ on 4F1 and ${\bf k_{i}}=2.02 \text{\AA}^{-1}$
on D7. The longitudinal XYZ-PA was performed using Helmoltz-like 
(4F1) and a quadrupolar assembly (D7) coils  \cite{fennell2017wavevector} allowing to choose the polarization
of the neutron either along \textbf{Z} always perpendicular to the
scattering plane. The \textbf{X} and \textbf{Y} polarizations correspond
to arbitrary directions within the scattering plane on D7, whereas
\textbf{X} is always set to be parallel to the scattering vector \textbf{Q}
on 4F1 and \textbf{Y} is perpendicular to \textbf{Q} but still within
the scattering plane. In all experiments, the samples were aligned
in the (1,0,0)/(0,0,1) scattering plane. The data are all reported
in reduced lattice units (r.l.u) and the measurement procedure follows
refs. \cite{mangin2015intra,tang2018orientation}.

On D7, the data were collected by performing $\pm10{^{\circ}}$ rocking trajectories around
the  (1,0,1) and (-1,0,0) positions.  We then performed data reduction
adapting the standard procedure \cite{stewart2009disordered}. Such
scans allowed the mapping of a wide \textbf{Q}-region spanning reflections
of the form (H,0,1) and satellite reflections (H,0,0.43) in r.l.u
of the ladders. Typical instrumental resolutions (full
width at half maximum: FWHM) are  0.07 r.l.u. and 0.03 r.l.u.  along the H and L directions, respectively.  
SF and NSF data were collected in the three channels
X,Y,Z. All data were corrected for the flipping ratio using a quartz
sample and the conversion to absolute units is done using a vanadium
sample. 

On 4F1, L and H-scans of the form (H,0,1) or (1,0,L) were done across reciprocal positions corresponding
to the ladder scattering ridge i.e. with integer L and H values.  Typical instrumental resolutions (FWHM) are  0.1 r.l.u. and 0.04 r.l.u. 
along the H and L directions, respectively.  Both SF and NSF scans were done in
order to crosscheck the absence of nuclear scattering at magnetic
positions. The PA across (H,0,1) was performed for positive and negative
H-values and the corresponding intensities were then symmetrized. Our experiments also include scans
along K. Such scans were performed by tilting the sample out of plane
using goniometers.


\bibliography{papier}

\begin{center}
\begin{figure}
\begin{centering}
\includegraphics[width=14cm]{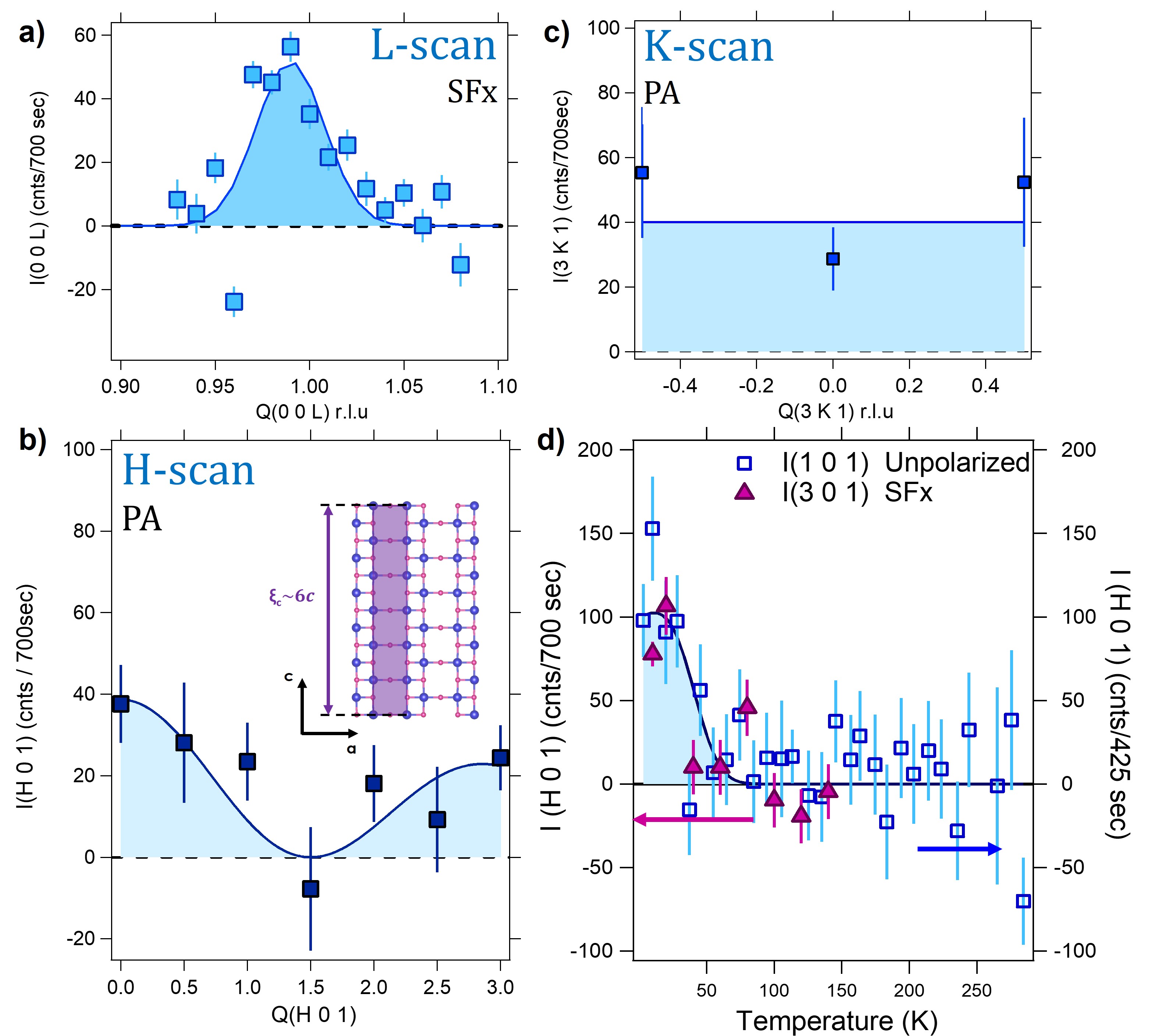} 
\par\end{centering}
\caption{\label{Fig3} \textbf{Magnetic scans in $Sr_{9}Ca_{5}Cu_{24}O_{41}$: (a)}
L-scan across (0,0,1) in the spin-flip ($SF_{X}$) channel, showing the magnetic
scattering along the ladders legs. The magnetic scattering appears
as a Gaussian signal centered at (0,0,1). (Raw data given in Supplementary
Note \textbf{2})\textbf{. (b)} H-scan of the (H,0,1) rod, the
inter-ladders direction (rungs), showing the modulation of the magnetic
intensity extracted from XYZ polarisation analysis (XYZ-PA) (squares).  The inset shows real-space
magnetic correlation lengths within the ladders planes. \textbf{(c)}
K-scan across (3,0,1) after full XYZ-PA, indicating the absence of
inter-plane magnetic correlations. \textbf{(d)} Temperature dependence
of the magnetic signal, after background subtraction (Raw data given
in Supplementary Note \textbf{3}). The figure combines polarized
neutron data (full symbols) measured at (3,0,1) in the $SF_{X}$ channel
and unpolarized neutron data (open symbols) measured at (1,0,1). All
\textbf{Q}-scans were performed on the instrument $4F1\text{ at }10K$. Lines are guide to the eye. Error bars represent one standard deviation. }
\end{figure}%
\par\end{center}

\begin{center}
\begin{figure}
\begin{centering}
\includegraphics[width=14cm]{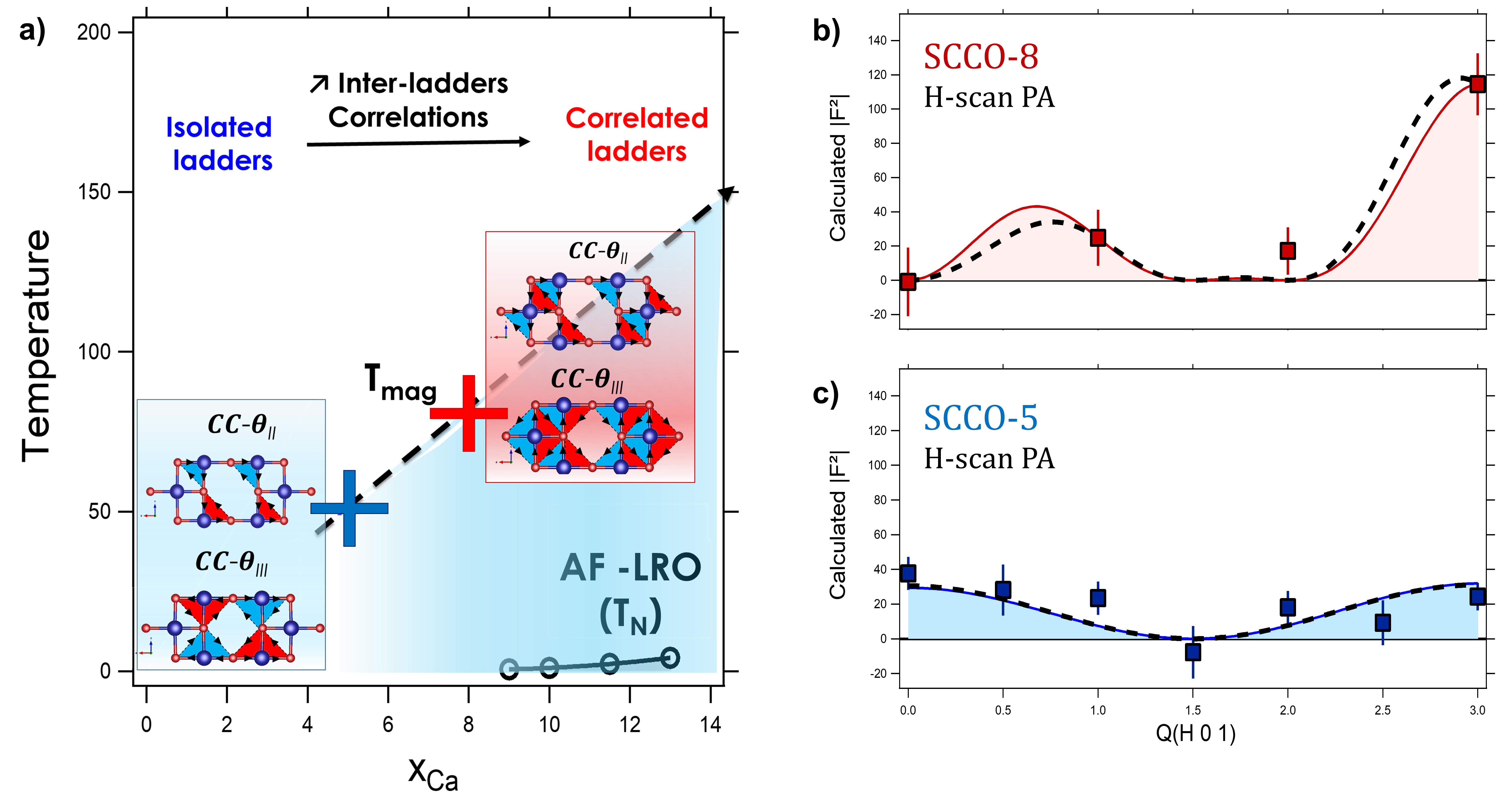} 
\par\end{centering}
\centering{}\caption{\label{Fig4}  \textbf{Phase diagram, modelisation and structure factors}:  
\textbf{(a)} Schematic phase diagram   in $Sr_{14-x}Ca_{x}Cu_{24}O_{41}$ ($SCCO-x$)
showing the evolution of the  loop current (LC) pattern as a function of the  $Ca$ content
($x_{Ca}$). 
At large doping, magnetic correlations develop between ladders at the
onset temperature $T_{mag}$ (red crosses).
In heavily doped samples, a magnetic long range order (LRO) further develops below a N\'eel
temperature $T_{N}$ of a few K \cite{vuletic2006spin}.
 Insets: \textbf{(A) } (top)  \textbf{$CC-\Theta_{II}$} \cite{simon2002detection,varma2006theory} and (down)  $CC-\Theta_{III}$ \cite{chatterjee2017,chatterjee2017intertwining,scheurer2018orbital} models  built on  single ladder with two staggered $Cu-O$ orbital currents per Cu site
flowing clockwise (red triangles) and anticlockwise (blue triangles). Both models described nicely the experimental results shown in panel  \textbf{(b)}. 
\textbf{(B)}   \textbf{$CC-\Theta_{II}$} and (down)  $CC-\Theta_{III}$ models with adddtional interladders correlations within the ladder unit cell. 
Both models described nicely the experimental results shown in panel  \textbf{(c)}.  \textbf{(b-c)} H-scan of the (H,0,1) rod, the inter-ladders direction
(rungs) showing the modulation of the out-of-plane magnetic intensity, $I_b$
extracted from XYZ polarisation analysis (PA) (squares) compared with magnetic structure factors
deduced from 2 loop currents models, labeled $CC-\Theta_{II}$ (solid
line) \cite{simon2002detection,varma2006theory} and $CC-\Theta_{III}$ (dashed line)\textbf{
}\cite{chatterjee2017,chatterjee2017intertwining,scheurer2018orbital} in \textbf{(b) $SCCO-5$ }and \textbf{(c)
}$SCCO-8$. Error bars represent one standard deviation. }
\end{figure}%
\par\end{center}

\begin{figure}
\begin{centering}
 \centering{} \includegraphics[height=11cm]{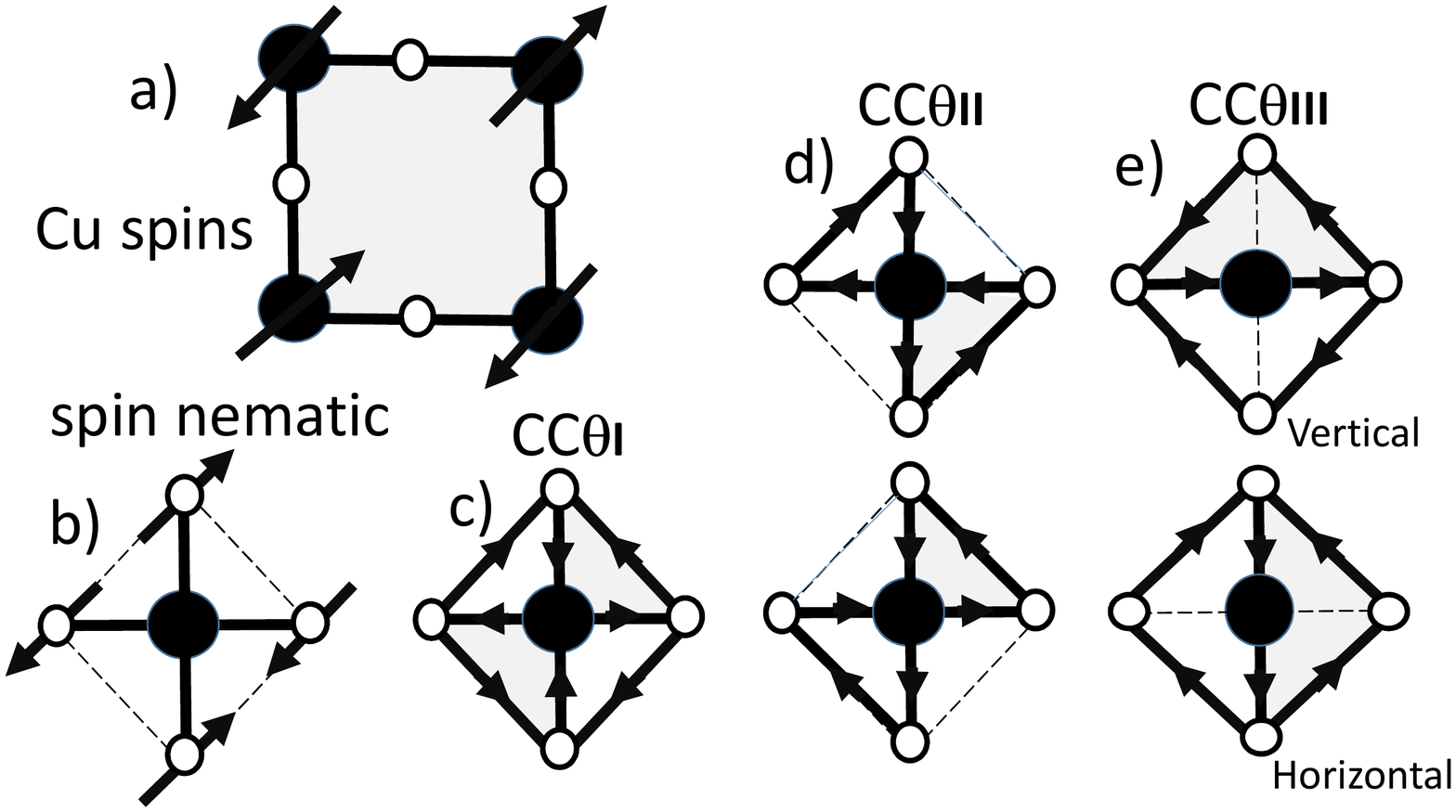}
\par\end{centering}
\caption{ \textbf{Magnetic and loop current (LC) patterns}:  (\textbf{a}) Periodic antiferromagnetic Cu spins on a square lattice. (\textbf{b}) a spin (or orbital) nematic state, with two sets
of staggered spin on O sites as proposed in \cite{fauque2006magnetic}. (\textbf{c}) LC state
$CC-\theta_{I}$ \cite{chudzinski2008orbital,chudzinski2010spin}.
(\textbf{d}) LC state $CC-\theta_{II}$ \cite{varma2006theory} showing two possible patterns breaking rotational symmetry along the diagonals. (\textbf{e})
LC state $CC-\theta_{III}$ \cite{scheurer2018orbital} showing \textit{Horizontal} and \textit{Vertical} patterns.
 For all models (b-e), the same pattern is assumed around each Cu atom on the square lattice of a single ladder.  }
\label{patterns} 
\end{figure}%

\clearpage

\section{Supplementary Material}
\beginsupplement


\section*{Supplementary material}

# 1 Supplementary Material

## SUPPLEMENTARY FIGURE 1: INSTRUMENT SETUP

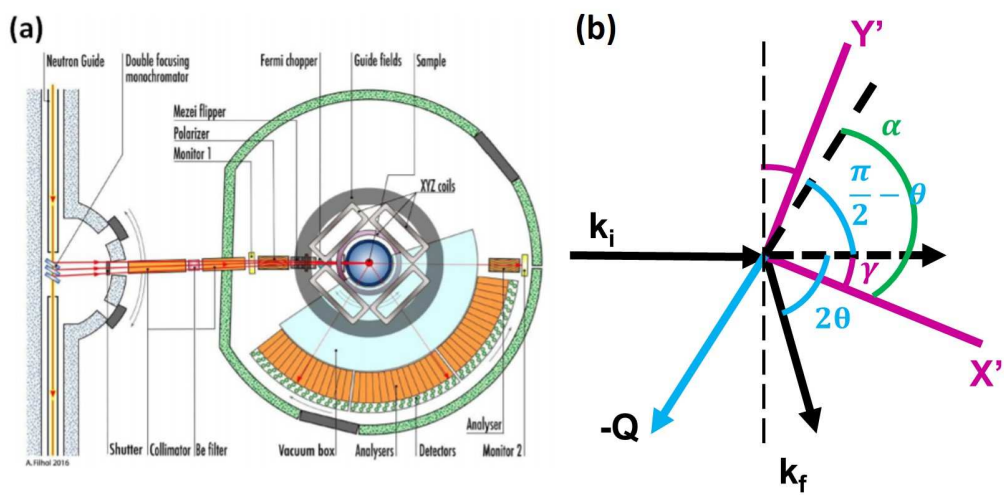

Supplementary Figure S1. (a) Layout of the multidetector diffractometer D7 [S1]. (b) Definition of the **X'** and **Y'** polarization directions within the scattering plane. $\gamma = 41.6°$ is the angle between the incident wave vector $\mathbf{k}_i$ and **X'**, set by the instrument configuration, $2\theta$ is the scattering angle, and $\alpha$ is defined as the angle between the wavevector **Q** and **X'**. Reproduced from [S2].

## SUPPLEMENTARY FIGURE 2: L-SCAN

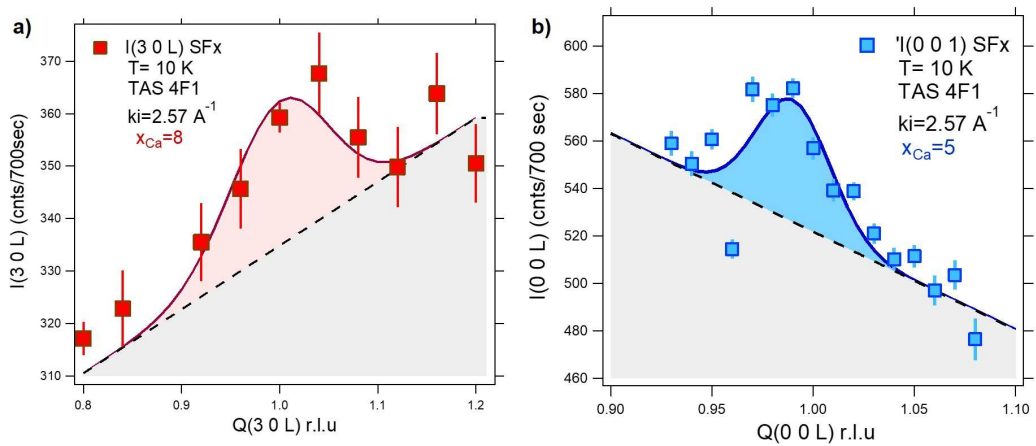

Supplementary Figure S2. L-scan measured at 10 K in the $SF_X$ channel. (**a**) around (3,0,1) for $SCCO-8$. (**b**) around (0,0,1) for $SCCO-5$. The scattered intensity is described by a Gaussian signal on top of the sloping background (shaded area). All measurements were performed on 4F1.

## SUPPLEMENTARY FIGURE 3: TEMPERATURE DEPENDENCE

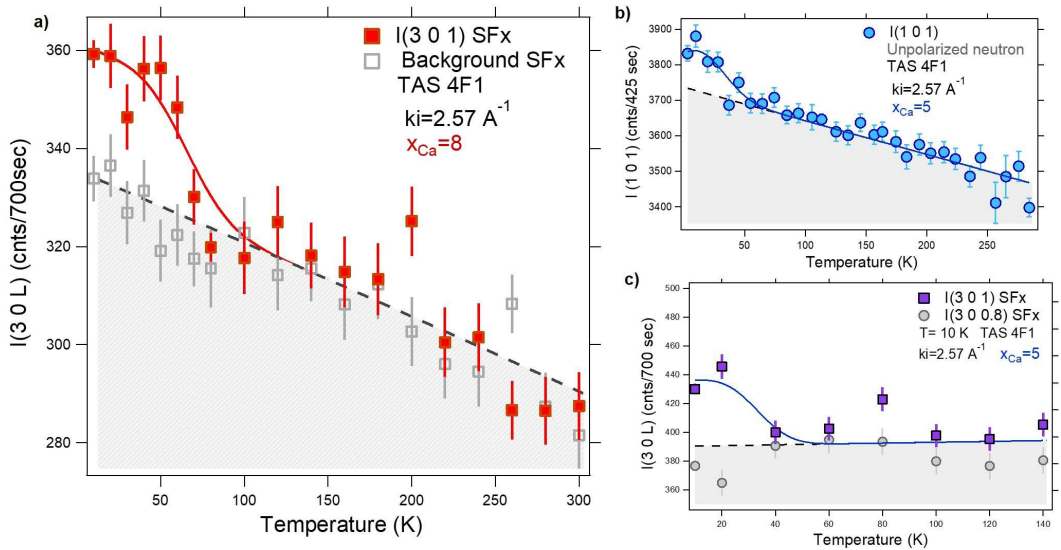

Supplementary Figure S3. (**a**) $SCCO-8$: T-dependence of the magnetic signal centered at (3,0,1) as measured in the $SF_X$ channel (red symbols), compared with the T-dependence of a background taken at positions shifted by $\Delta L = \pm 0.2$ away from (3,0,1) (see text). (**b-c**) $SCCO-5$: (**b**) Unpolarized neutron measurements of the T-dependence of the scattered intensity at (1,0,1) (blue circles). The departure from a linear T-dependence is shown by the dashed line. The shaded area allows the identification of the background to which an extra magnetic scattering adds at low temperature. (**c**) T-dependence of the magnetic signal at (3,0,1) as measured in the $SF_X$ channel (violet squares) compared to the T-dependence at the background position (3 0 0.8) (gray circles). All measurements were performed on 4F1.

## SUPPLEMENTARY FIGURE 4: ABSENCE OF THE MAGNETIC INTENSITY WITHIN THE CHAIN SUBSYSTEM

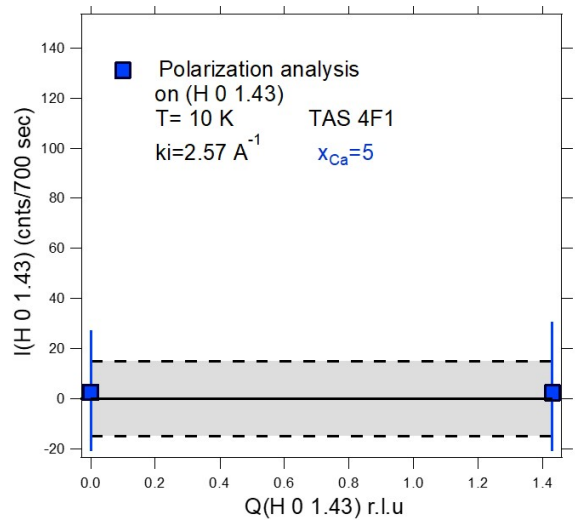

Supplementary Figure S4. Full magnetic intensity deduced from XYZ-PA (4F1, T=10K) within the chain subsystem along (H,0,0,1) in superspace r.l.u or (H,0,1.43) in ladders r.l.u for $SCCO-5$.

## SUPPLEMENTARY FIGURE 5: IN-PLANE AND OUT-OF-PLANE MAGNETIC INTENSITY

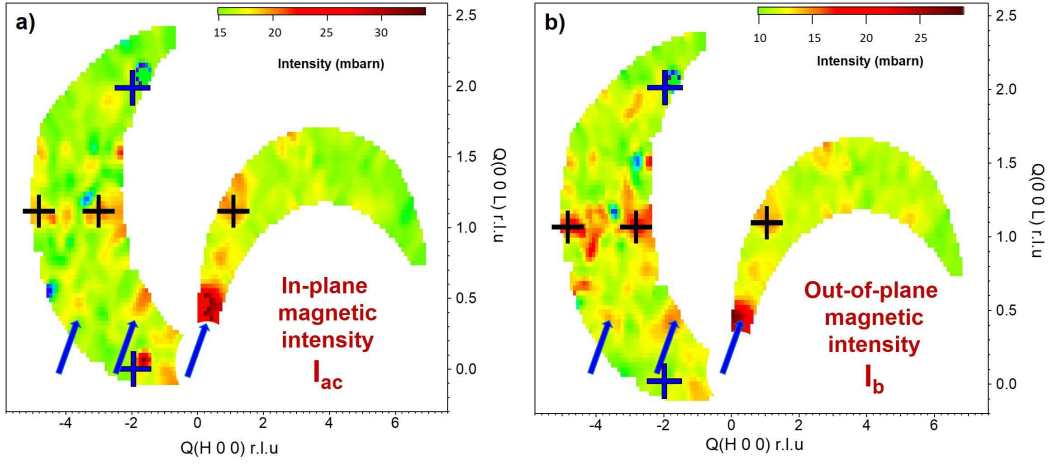

Supplementary Figure S5. $SCCO-8$ : Mapping of the magnetic intensities within ladders at T=5 K, as deduced from XYZ-PA carried out using SF data measured on $D7$: (**a**) the in-plane intensity $I_{ac}$, (**b**) the out-of-plane intensity $I_b$. The maps are given in r.l.u of the ladder subsystem and the intensities calibrated in $mbarn$. The magnetic spots are located by crosses along the ladder scattering ridge along (H,0,1). The blue arrows show the satellite magnetic reflections at (H,0,1,-1) using superspace notations or L∼0.43 in ladders r.l.u. The blue crosses indicate the positions of Nuclear Bragg scattering from the ladders. Red spots around these two regions do not correspond to magnetic scattering but are due to polarization leakage.

## SUPPLEMENTARY FIGURE 6: ABSENCE OF CDW

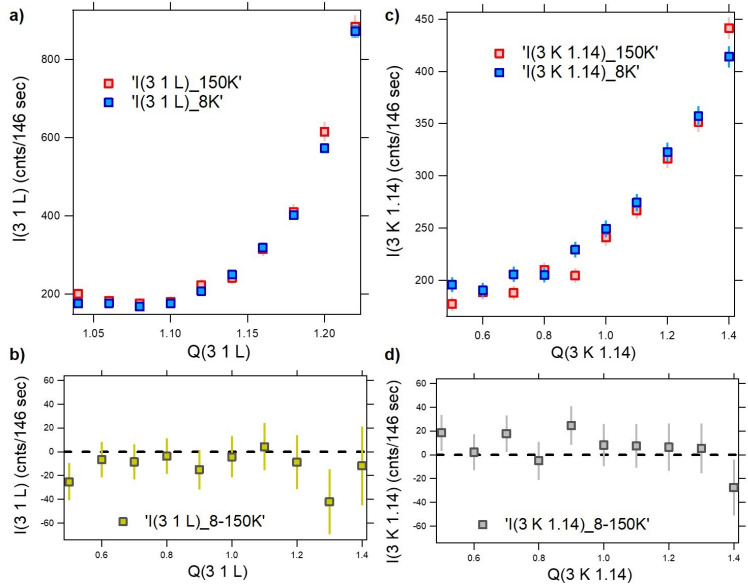

Supplementary Figure S6. **$SCCO-5$**: series of scans performed on 4F1 in the $NSF_X$ channel at 8 K (red symbols) and 150 K (blue symbols). (**a**) scans along (3,1,L) crossing both L=1.2 and L=1.14 positions in r.l.u of the ladders, corresponding to $q_{CDW}$ reported for ladders and chains, respectively. (**c**) K-scans across (3,K,1.14) in r.l.u of the ladders, corresponding to $q_{CDW}$ reported for chains. The increase of the intensity in both large $L$ or $K$ is due scattering of Al from the sample holder. (**b,d**) Differential intensities 8 K-150 K, from scans reported in (**a,c**) that show the absence of any $CDW$-induced structural distortion.

## SUPPLEMENTARY FIGURE 7: MAGNETIC PATTERNS

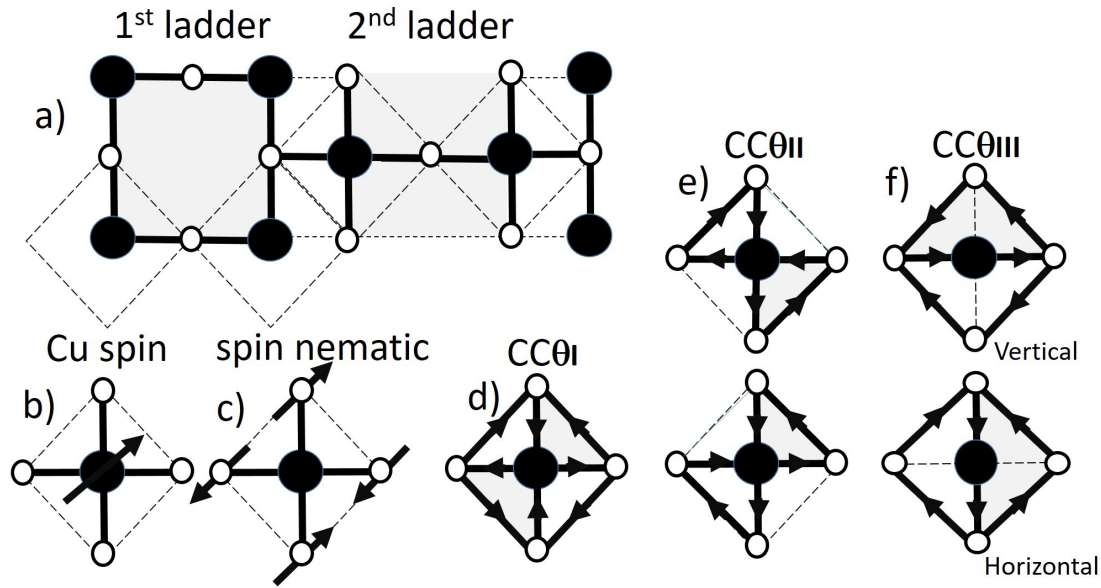

Supplementary Figure S7. (**a**) Ladder unit cell in SCCO containing Copper sites (full dots) and Oxygen sites (open dots). (**b-f**) The $CuO_2$ plaquette can be decorated with various magnetic patterns: (**b**) a spin on Cu site. (**c**) a spin (or orbital) nematic state, with two sets of staggered spin on O sites as proposed in [S3]. (**d**) loop current (LC) state $CC - \theta_I$ [S4, S5]. (**e**) LC state $CC - \theta_{II}$ [S6] showing two possible patterns breaking rotational symmetry along the diagonals: upper part denoted $\epsilon = +1$ and lower part denoted $\epsilon = -1$. (**f**) LC state $CC - \theta_{III}$ [S7]: lower part (*Horizontal*) and upper part (*vertical*) pattern, respectively. For all LC states, the staggered orbital moments are expected to point perpendicular to the circulating currents (ladder plane).

## SUPPLEMENTARY FIGURE 8: ANTIFERROMAGNETIC MODEL

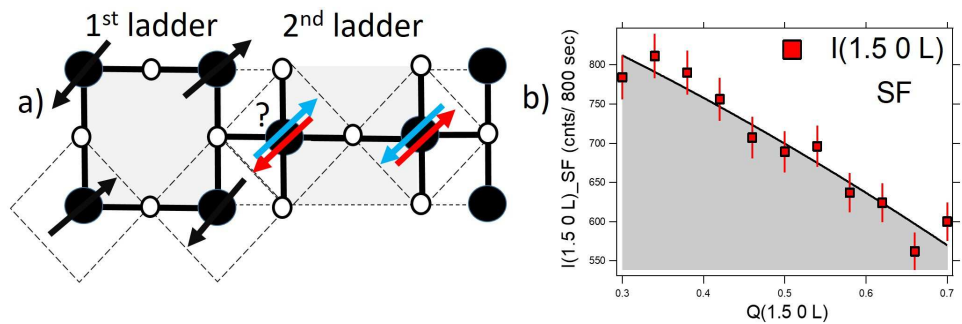

Supplementary Figure S8. (**a**) Model of antiferromagnetically interacting Cu spins in the first ladder. The question mark and colored arrows show the frustrated Cu spin on the second ladder, suggesting a high degree of frustration between neighboring ladders. (**b**) L-scan across (1.5,0,L) as measured on 4F1 at 12 K in the $SF_X$ channel.

### SUPPLEMENTARY FIGURE 9: MAGNETIC MOMENTS ON THE OXYGEN SITES

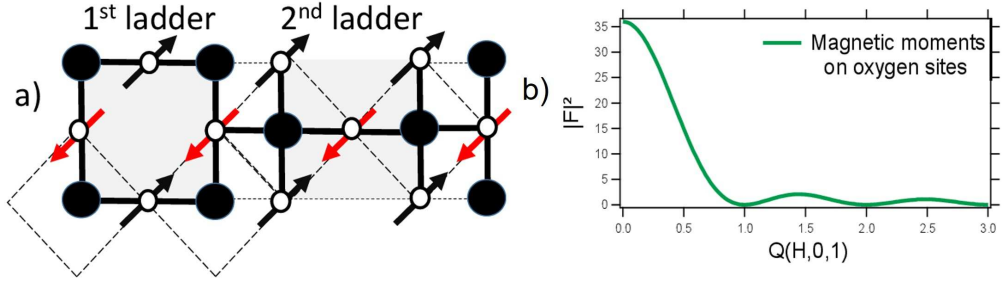

Supplementary Figure S9. (**a**) Magnetic nematic order, with staggered magnetic moments on O sites in each $CuO_2$ plaquette. **b**, H-dependence of $|F(\mathbf{Q})|^2$ for a model of magnetic moments on oxygen sites for L=1

### SUPPLEMENTARY FIGURE 10: Q-DEPENDENCE OF LC PATTERNS

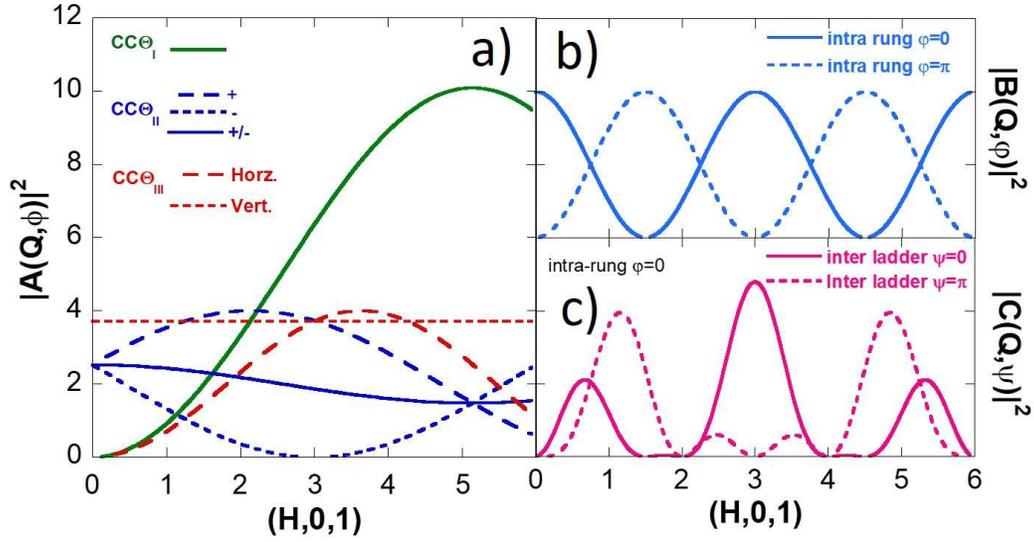

Supplementary Figure S10. Q-dependence along (H,0,1): (**a**) the squared magnetic structure factor $|A_\phi(\mathbf{Q})|^2$ associated with each of the LC patterns: $CC-\theta_I$, $CC-\theta_{II}$ ($+$, $-$ and average (see text) denoted $+/-$) and $CC-\theta_{III}$ (*Horizontal* and *Vertical*). (**b**) The structure factor $|B_{\phi\phi'}(\mathbf{Q})|^2$ coming from the intra-rung correlations: $\varphi = 0$ for identical patterns and $\varphi = \pi$ for opposite patterns. (**c**) The structure factor $|C_{\phi\phi'}(\mathbf{Q}))|^2$ coming from the inter-ladder correlations: $\psi = 0$ for identical patterns and $\psi = \pi$ for opposite patterns. (The intra-rung term, $|B_\phi(\mathbf{Q})|^2$ is taken for $\varphi = 0$).

**SUPPLEMENTARY FIGURE 11: $CC - \Theta_I$ LC PATTERNS**

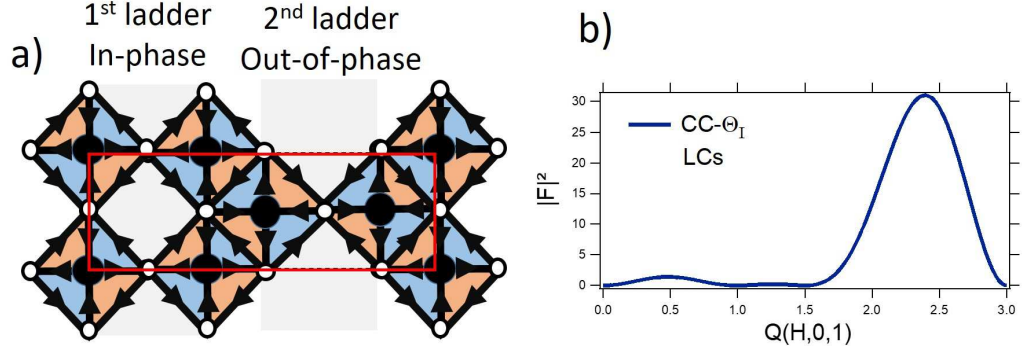

Supplementary Figure S11. (**a**) $CC - \Theta_I$ state [S6]: 4 LCs per $CuO_2$ flowing clockwise (blue triangles) and anticlockwise (red triangles). (b) H-dependence of the structure factor for a model of $CC - \theta_I$ LCs for L=1.

**SUPPLEMENTARY FIGURE 12: $CC - \Theta_{II}$ LC PATTERNS**

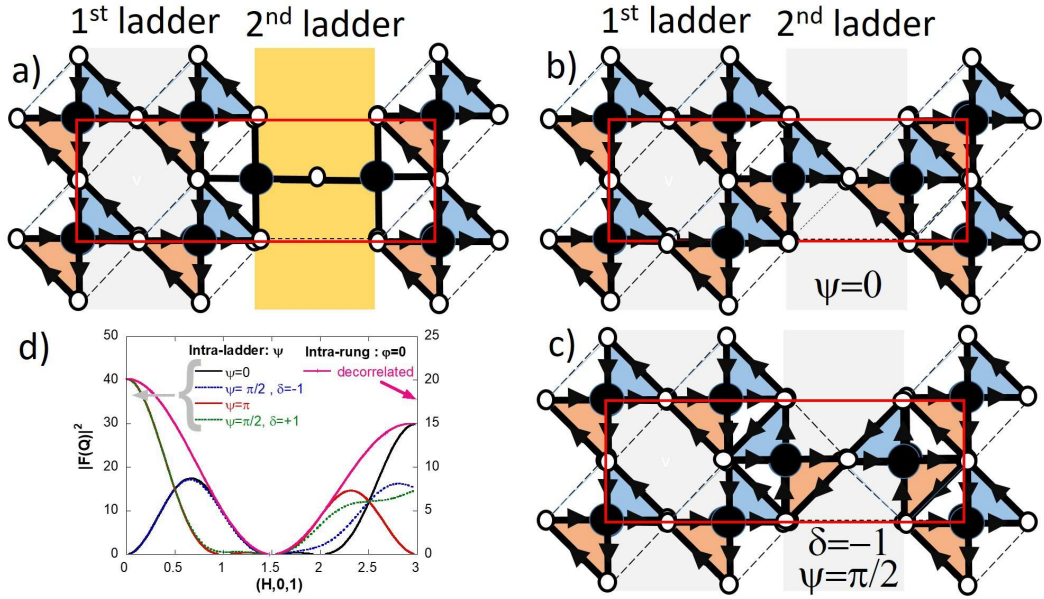

Supplementary Figure S12. $CC - \Theta_{II}$ state [S6]: 2 LCs per $CuO_2$ flowing clockwise (blue triangles) and anticlockwise (red triangles) and aligned along a given diagonal. (**a**) Uncorrelated model: only 2 two Cu sites in the first ladder are decorated with the same $CC - \Theta_{II}$ pattern, whereas the second ladder remains non magnetic. (**b**) Correlated model : 2 two Cu sites in the first ladder and the two other in the second one are decorated with the same $CC - \Theta_{II}$ pattern (case : in-phase, $\varphi = 0$). (**c**) Correlated model: 2 two Cu sites in the first ladder exhibit the same LC patterns, while the two other in the second one are characterized by a LC pattern rotated at $\frac{\pi}{2}$ (case : crisscrossed, $\psi = \frac{\pi}{2}$, $\delta = -1$). (**d**) $|F(\mathbf{Q})|^2$ for the uncorrelated and in-phase case (magenta) and the correlated one, considering 4 distinct cases: in-phase, $\psi = 0$ (black), crisscrossed, $\psi = \pm\frac{\pi}{2}$, $\delta = +1$ green), out-of-phase, $\psi = \pi$ (red), crisscrossed, $\psi = \pm\frac{\pi}{2}$, $\delta = -1$ (blue).

## SUPPLEMENTARY FIGURE 13: $CC - \Theta_{III}$ LC PATTERNS

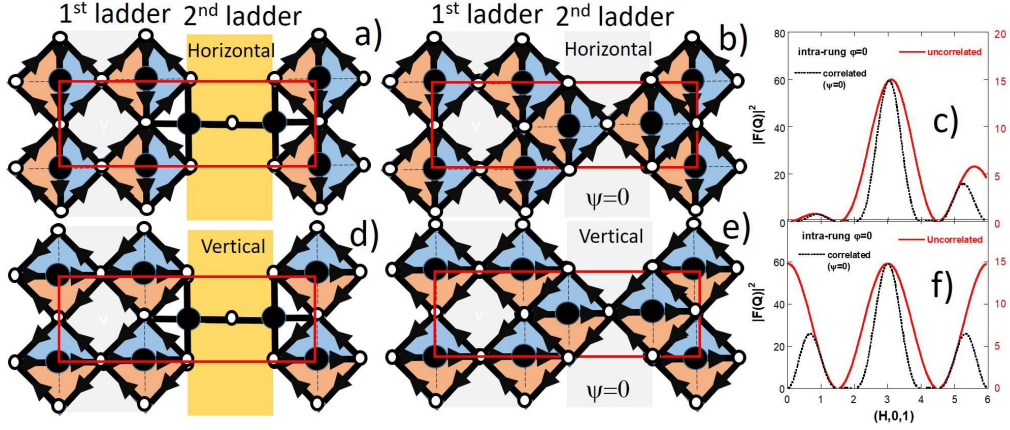

Supplementary Figure S13. $CC - \Theta_{III}$ state [S7, S8]: 2 LCs per $CuO_2$ flowing clockwise (blue triangles) and anticlockwise (red triangles) and aligned along either the ladder rungs (*Horizontal*, (**a-c**)) or the ladder legs (*Vertical*, (**d-f**)) . (**a,d**) Uncorrelated model (in-phase intra-rung coupling: $\varphi$=0): Only 2 two Cu sites in the first ladder are decorated with the same $CC - \Theta_{III}$ pattern; whereas the second ladder remains non magnetic. (**b,e**) Correlated model (in-phase inter-ladder coupling : $\psi$=0): 2 two Cu sites in the first ladder and the two other in the second one are decorated with the same $CC - \Theta_{III}$ pattern. (**c,f**) Squared magnetic structure factors: uncorrelated case (red solid line) and correlated case (black dotted line).

## SUPPLEMENTARY FIGURE 14: COMPARISON OF MEASURED INTENSITIES AND LC MODELS

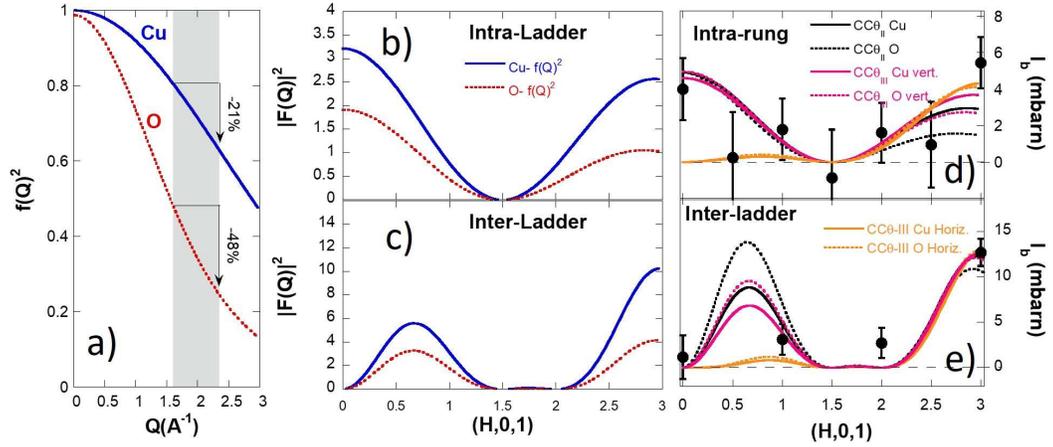

Supplementary Figure S14. (**a**) **Q**-dependencies of the squared magnetic form factor for oxygen $f_O^2(Q)$ and copper $f_{Cu}^2(Q)$. The shaded area indicates the Q-range where the PND study was carried out. (**b,c**) Q-dependencies of $(f(Q)^2 |F(\mathbf{Q})|^2)$ along (H,0,1): (**b**) uncorrelated case with in-phase intra-rung correlation only, (**c**) correlated case, in-phase inter-ladder correlation. (**d,e**) Q-dependencies of of $I_b$ (mbarn) along (H,0,1): (**d**) uncorrelated case with in-phase intra-rung correlation only, (**e**) correlated case, in-phase inter-ladder correlation. The fits correspond to different LC patterns: $CC - \theta_{II}$ (black), $CC - \theta_{III}$ *Vertical* (magenta) and *Horizontal* (orange). The solid and dotted lines correspond to fits including the Cu- and O-form factors, respectively.

## SUPPLEMENTARY FIGURE 15: Q-DEPENDENCIES OF THE ORIENTATION FACTORS

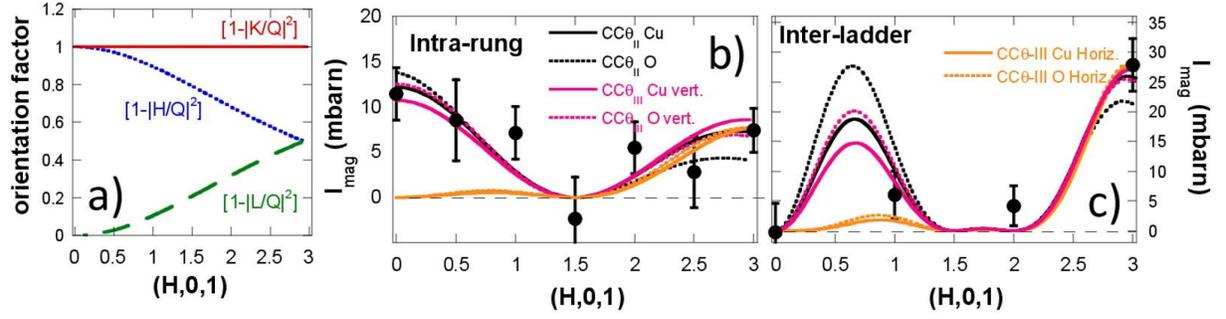

Supplementary Figure S15. (**a**) **Q**-dependencies of the orientation factors $[1 - \frac{Q_i^2}{Q^2}]$, with $i = \{a, b, c\}$. (**b,c**) **Q**-dependencies of $I_{mag}$ along (H,0,1): (**b**) SCCO-5 data, with fits for the uncorrelated case with in-phase intra-rung correlations only, (**c**) SCCO-8 data with fits for the the correlated case with in-phase intra-ladder correlations. The fits correspond to different LC patterns: $CC - \theta_{II}$ (black), $CC - \theta_{III}$ *Vertical* (magenta) and *Horizontal* (orange). The solid and dotted lines are associated with fits including the Cu- and O-form factors, receptively.

## SUPPLEMENTARY NOTE 1: POLARIZED NEUTRON DIFFRACTION (PND) AND XYZ-POLARIZATION ANALYSIS (XYZ-PA)

### Neutron diffraction

The PND experiments were carried out on two instruments: the 4F1 triple axis spectrometer (Orphée reactor, Laboratoire Léon Brillouin, Saclay) and the D7 multidetector diffractometer (Institut Laue Langevin, Grenoble). On each instrument, measurements were carried out with a neutron wavevector of 2.57 Å$^{-1}$ and 2.02 Å$^{-1}$, respectively.

$Sr_{14-x}Ca_xCu_{24}O_{41}$ ($SCCO$-$x$) exhibits an aperiodic atomic structure with two, ladders (ld) and chains (ch), incommensurate sub-lattices. Therefore, Bragg peaks need to be indexed in the 4D superspace as $(H, K, L_{ld}, L_{ch})$. Both sublattices are incommensurate along the **c**-axis with an incommensurability parameter $\frac{1}{\gamma} = \frac{c_{ch}}{c_{ld}} = 1.42$. In the experiments, we used the ladder sub-lattice parameters as a reference: H, K and L coordinates are expressed in reduced lattice units (r.l.u) of the ladder (ld) subsystem: $\frac{2\pi}{a} = 0.55$ Å$^{-1}$, $\frac{2\pi}{b} = 0.48$ Å$^{-1}$ and $\frac{2\pi}{c_{ld}} = 1.61$ Å$^{-1}$, respectively.

The samples were always aligned in the (H,0,0)/(0,0,L) scattering plane, in order to probe the [a,c] plane of the ladders. Within that scattering plane, a transferred momentum **Q** of the form (H,0,L) is accessible. Within these units, the Bragg peaks associated with the chains subsystem contribute at $L_{chains} = 1.42L$ as $\frac{2\pi}{c_{ch}} = 2.28$ Å$^{-1}$. Additionally, tilting the sample out of the scattering plane using goniometers allowed us to access wavevectors of the form **Q**(H,K,L) on 4F1.

Considering a magnetic sample, the magnetic scattering cross-section [S9], $I_{mag}$, reads:

$$I_{mag} = \Phi_S \, N_{cell} \, r_0^2 \, f(Q)^2 \, |F(\mathbf{Q})|^2 \, m_\perp^2 \tag{1}$$

$\Phi_S$ corresponds to the neutron flux at the sample in $n/s/barns$. $N_{cell}$ is the number of unit cells in the sample. $r_0$ stands for the neutron magnetic scattering length, $r_0^2 = 290 \, mbarn$. $f(Q)$ is to the magnetic form factor and $|F(\mathbf{Q})|$ the magnetic structure factor. Owing to the dipolar nature of the interaction between neutron spin and the magnetic moments, **m**, $I_{mag}$ probes $\mathbf{m}_\perp$ the magnetic components perpendicular to **Q**, only.

The squared modulus of $\mathbf{m}_\perp$ can be expressed using the regular Cartesian coordinates of the lattice:

$$m_\perp^2 = \mathbf{m}^2 - (\mathbf{m}.\mathbf{Q})^2 = \sum_{i,j=a,b,c} \left(\delta_{ij} - \frac{Q_i Q_j}{Q^2}\right) m_i m_j \tag{2}$$

For a magnetic moment, $\mathbf{m}(\pm m_a, \pm m_b, \pm m_c)$ with $n$ non zero components, there are $2^n$ equivalent magnetic domains. The cross-terms ($i \neq j$) cancel out when summing over all domains, at variance with the diagonal terms ($i = j$). In the case of the present study with **Q**=(H,0,L), $I_{mag}$ reduces to the sum of in- and out-of plane terms, so that:

$$I_{mag} = I_{ac} + I_b \tag{3}$$

With $I_{ac} \propto m_{ac}^2 = [(1 - |\frac{H}{Q}|^2)m_a^2 + (1 - |\frac{L}{Q}|^2)m_c^2]$ and $I_b \propto m_b^2$ (because $K = 0$).

It is also convenient to use the so-called {XYZ} referential, where **X** is the unitary vector parallel to **Q**. **Y** and **Z** are the two unitary vectors orthogonal to **X**, within the scattering plane and perpendicular to the scattering plane, respectively. So that $\mathbf{m}_\perp = m_Y \mathbf{Y} + m_Z \mathbf{Z}$, with the in- and out-of (scattering) plane components.

### Polarized neutron diffraction setup

On the incoming neutron beam, a bender (polarizing super-mirror) can polarize the neutron spin and a Mezei flipper can adiabatically flip the neutron spin. A pyrolitic graphite filter is further added before the bender to remove high harmonics on 4F1. The neutron spin polarization is maintained using a homogeneous guide field of a few Gauss. The neutron spin polarization direction, **P** is controlled on the sample by Helmholtz coils on 4F1 and a quadrupolar assembly on D7 [S1]. On the scattered neutron beam, the final neutron spin polarization is analyzed using either an analyzer made of co-aligned Heusler single crystals (on 4F1) or polarizing benders (on D7) placed in front of the multidetector bank (Supplementary Figure S1.a)

**Polarized neutron diffraction cross sections**

Polarization analysis of PND allows us to distinguish between the different contributions to the scattered intensity [S9]. For a nuclear scattering, the neutron spin remains unchanged and the scattered intensity is measured in the Non Spin-Flip (NSF) channel. Since the spin polarization of the neutron beam is not perfect, a small amount of the nuclear scattering can nevertheless leak into the Spin-Flip (SF) channel. The ratio between scattered intensities in the $NSF$ and the $SF$ channels is called the flipping ratio $FR$ and characterizes the quality of spin polarization of the neutron beam. For a magnetic scattering, the scattered intensities in each channel strongly depends on $\mathbf{P}$. Indeed, neutron spins are described using Pauli matrices, whose quantization axis is given by $\mathbf{P}$. The magnetic intensity $I_{mag}^{NSF}(\mathbf{P}) \propto (\mathbf{m}_\perp.\mathbf{P})^2$ does not flip the neutron spin and remains in the NSF channel. The remaining magnetic $I_{mag}^{SF}(\mathbf{P}) \propto (\mathbf{m}_\perp)^2 - (\mathbf{m}_\perp.\mathbf{P})^2$ appears in the SF channel.

On D7, the direction of polarization $\mathbf{X}$ is not parallel to $\mathbf{Q}$ but along a direction $\mathbf{X}'$ turned by an angle $\alpha$ (Supplementary Figure S1.b). Therefore, one needs to estimate the scattered intensities in directions of $\mathbf{P}$ corresponding to unitary vectors $(\mathbf{X}', \mathbf{Y}', \mathbf{Z}')$, as shown in Supplementary Figure S1.b, given by:

$$\begin{pmatrix} \mathbf{X}' \\ \mathbf{Y}' \\ \mathbf{Z}' \end{pmatrix} = \begin{pmatrix} cos\alpha & sin\alpha & 0 \\ -sin\alpha & cos\alpha & 0 \\ 0 & 0 & 1 \end{pmatrix} \begin{pmatrix} \mathbf{X} \\ \mathbf{Y} \\ \mathbf{Z} \end{pmatrix} \quad (4)$$

Then, the full magnetic scattering, $I_{mag}$, given in Eq. 3, splits in two terms, $I_{mag}^{SF}(\mathbf{P})$ and $I_{mag}^{NSF}(\mathbf{P})$

$$\begin{cases} I_{mag}^{SF}(\mathbf{X}') = I_{ac}\, cos^2\alpha + I_b & I_{mag}^{NSF}(\mathbf{X}') = I_{ac}\, sin^2\alpha \\ \\ I_{mag}^{SF}(\mathbf{Y}') = I_{ac}\, sin^2\alpha + I_b & I_{mag}^{NSF}(\mathbf{Y}') = I_{ac}\, cos^2\alpha \\ \\ I_{mag}^{SF}(\mathbf{Z}') = I_{ac} & I_{mag}^{NSF}(\mathbf{Z}') = I_b \end{cases} \quad (5)$$

In addition to the polarization magnetic cross-sections of Eq. 5, $I_{mag}(\mathbf{P})$, one should consider the nuclear intensity, $I_{nucl}$ and a background, $Bg$, in both SF and NSF channels. Both SF and NSF cross-sections read:

$$\begin{cases} I^{NSF}(\mathbf{P}) \sim Bg^{NSF} + I_{nucl} + I_{mag}^{NSF}(\mathbf{P}) \\ \\ I^{SF}(\mathbf{P}) \sim Bg^{SF} + I_{mag}^{SF}(\mathbf{P}) \end{cases} \quad (6)$$

As discussed in [S10], due to imperfect polarizations, the measured neutron intensities are mixing the cross-sections of Eq. 6. In each channel, it can be actually written,

$$\begin{cases} I_{meas}^{NSF}(\mathbf{P}) & \propto Bg^{NSF} + I_{nucl} + I_{mag}^{NSF}(\mathbf{P}) + \frac{1}{FR(\mathbf{P})}[Bg^{SF} + I_{mag}^{SF}(\mathbf{P})] \\ \\ I_{meas}^{SF}(\mathbf{P}) & \propto Bg^{SF} + I_{mag}^{SF}(\mathbf{P}) + \frac{1}{FR(\mathbf{P})}[Bg^{NSF} + I_{nucl} + I_{mag}^{NSF}(\mathbf{P})] \end{cases} \quad (7)$$

where $FR(\mathbf{P})$ is the polarization dependent flipping ratio of the experiment. For both instruments, 4F1 and D7, $FR(\mathbf{P})$ were measured for all relevant scattering angle using a quartz sample. One can deduce the PND cross-sections, Eq. 6, from the measured ones, Eq. 7 [S10]. In the present study, although the nuclear and magnetic scattering preserve the lattice translation symmetry, the short range magnetism (SRM) occurs at $\mathbf{Q}$ values where there is additional extinction of the nuclear peaks due to the 3D atomic structure. Therefore, the effects of imperfect polarizations of the nuclear term and background terms of Eq. 7 are relatively weak (although sizeable) and readily corrected. It should be emphasized that this is a very different situation from the case of most superconducting cuprates [S2, S3] where the q=0 magnetism occurs at the same Bragg position of the nuclear structure.

**XYZ-Polarization analysis**

We systematically performed a longitudinal XYZ polarization analysis (XYZ-PA), that allows the determination of the full magnetic intensity $I_{mag} = I_{ac} + I_b$ ( Eq. 3) from a set of measurements in the SF channel with **P** along each of the 3 unitary vectors $(\mathbf{X'}, \mathbf{Y'}, \mathbf{Z'})$. The combination of all those measurements provides access to the in- (// ac) and out-of-(// b) scattering plane magnetic scattering and the SF background. Two different situations occur for both instruments:

• On *D7*, Diffractometer: $\alpha = 90 - \theta + 41.6°$ (Supplementary Figure S1), with $\theta$ the Bragg angle that depends on **Q** and the neutron wave length. From **X'**, **Y'** and **Z'** measurements, the full measurable magnetic intensity is deduced from Eq. 5 as follows :

$$\begin{cases} I_{ac} = & [I^{SF}(\mathbf{X'}) - I^{SF}(\mathbf{Y'})]/[cos^2\alpha - sin^2\alpha] \\ \\ I_b = & [I^{SF}(\mathbf{X'}) - I^{SF}(\mathbf{Z'})] + sin^2\alpha \, I_{ac} \\ \\ I_{mag} = & [I^{SF}(\mathbf{X'}) - I^{SF}(\mathbf{Z'})] + g(\alpha)[I^{SF}(\mathbf{X'}) - I^{SF}(\mathbf{Y'})] \end{cases} \quad (8)$$

For $I_{mag}$, the second term is weighted by $g(\alpha) = [1 + sin^2\alpha]/[cos^2\alpha - sin^2\alpha]$, which goes to unity for $\alpha$=0.

• On *4F1*, Triple axis spectrometer: $\alpha = 0$, one obtains the usual relations:

$$\begin{cases} I_{ac} = & I^{SF}(\mathbf{X}) - I^{SF}(\mathbf{Y}) \\ \\ I_b = & I^{SF}(\mathbf{X}) - I^{SF}(\mathbf{Z}) \\ \\ I_{mag} = & 2\,I^{SF}(\mathbf{X}) - I^{SF}(\mathbf{Y}) - I^{SF}(\mathbf{Z}) \end{cases} \quad (9)$$

**SUPPLEMENTARY NOTE 2: L-DEPENDENCE OF THE SHORT RANGE MAGNETISM (SRM)**

Supplementary Figure S2.a shows the SF intensity measured along (3,0,L) for $\mathbf{P}//\mathbf{X}$ at 10 K on sample $SCCO-8$. On the L-scan, the magnetic signal at L=3, exhibits a Gaussian line-shape and appears on top of a sloping background. The determination of such a sloping background has been confirmed by XYZ-PA. A qualitatively similar type of signal is observed for sample $SCCO-5$ (Supplementary Figure S2.b).

**SUPPLEMENTARY NOTE 3: TEMPERATURE DEPENDENCE OF THE SHORT RANGE MAGNETISM**

• $SCCO-8$: the T-dependence of the scattered intensity at (3,0,1) was measured in the $SF_X$ channel on 4F1 (Supplementary Figure S3.a). It displays a net enhancement at low temperature. According to the L-scan across (3,0,1) performed at low temperature (Fig. 2.b in the main text), the SRM signal centered at L=1 vanishes at L=0.8 and 1.2. Two additional T-dependencies were measured at those L values and averaged to provide the T-dependence of the non magnetic background. The comparison of both T-dependencies indicates that the magnetic signal starts developing below an onset temperature $T_{mag} \simeq$80 K.

• $SCCO-5$: The raw temperature dependencies of the magnetic scattering at (1,0,1) and (3,0,1), were measured on 4F1 (Supplementary Figure S3.b-c). Using an unpolarized neutron beam, the scattered intensity at (1,0,1) exhibits a linear increase on cooling down to a temperature where an extra enhancement of the intensity becomes visible. Using a polarized neutron beam, the signal at (3,0,1) was measured in $SF_X$ and compared to the scattered intensity at (3,0,0.8), a background position according to measurements in $SCCO-8$. The comparison between polarized and unpolarized neutron data highlights an onset temperature $T_{mag} \simeq$50 K below which the SRM sets in. XYZ-PA has been also employed at $\sim$ 100K at (0,0,1), showing the vanishing of the SRM at high temperature.

**SUPPLEMENTARY NOTE 4: ABSENCE OF MAGNETIC SIGNAL WITHIN THE CHAINS**

A new magnetic signal is clearly observed at $(H, 0, L)$ with integer values of $H$ and $L$, corresponding to the ladder subsystem. However, no magnetic signal is observed for the chain subsystem at Bragg positions $(H, 0, 0, 1)$ using superspace notations, corresponding to $(H, 0, 1.43)$ in the ladder subsystem units. A scan performed across $(H, 0, 1.43)$, shown in Supplementary Figure S4, shows the full scattered magnetic intensity as deduced from XYZ-PA (4F1) in $SCCO-5$ (which complements a similar scan in $SCCO-8$ in the Fig. 3.a of the main text). In both samples, the XYZ-PA reveals the absence of scattered magnetic intensity at positions corresponding to the chains subsystem.

**SUPPLEMENTARY NOTE 5: CALIBRATION IN ABSOLUTE UNITS**

We converted the measured intensities in absolute units using a vanadium sample, which allows the determination of the neutron flux at the sample position $\Phi_S$. A vanadium sample is a pure incoherent scatterer. For PND measurements, $2/3$ of its intensity shows up in the SF channel ($1/3$ in the NSF channel) and its energy ($\omega$) integrated intensity reads:

$$I_{Vana}^{SF} = \Phi_S . N_{cell} . \frac{2}{3} (\frac{d\sigma}{d\Omega})^{inc} \tag{10}$$

Where $(\frac{d\sigma}{d\Omega})^{inc} = 0.394\, barns$ stands for the vanadium incoherent cross section. For a vanadium sample mass, $m_V = 1\, g$, and a molar mass, $M_V = 50.94\, g.mol^{-1}$, one obtains $N_{Cell} = 0.6023.\frac{m_V}{M_V} = 0.0118\, cells/mol$.

The incoherent scattering for vanadium is purely elastic and is described by a Dirac distribution in energy, $\delta(\omega)$. The measured $\omega$-dependence is obtained after convolution by the Gaussian instrumental energy resolution, characterized by a full width at half maximum (FWHM) $\Delta_\omega$: the measured intensity $I_{meas}(\omega)$ acquires a Gaussian profile as,

$$I_{meas}(\omega) = I_{max}\, exp[-4\, ln(2) \frac{\omega^2}{\Delta_\omega^2}] \tag{11}$$

Integrating over energy, one obtains: $I_{Vana}^{SF} = \frac{I_{max}\Delta_\omega}{2} \sqrt{\frac{\pi}{ln(2)}}$.

Taking into account the instrument energy resolution : $\Delta_\omega = 1.25\, meV$ for $k_i = k_f = 2.57\, \text{Å}^{-1}$ gives $\Phi_S = 1678\, n/s/barns$. This value of $\Phi_S$ holds for both experiments on $SCCO-5$ and $SCCO-8$ on the spectrometer 4F1. A similar procedure using a vanadium standard sample was used for data calibration on D7 [S10].

**SUPPLEMENTARY NOTE 6: IN-PLANE AND OUT-OF-PLANE MAGNETIC SCATTERING AMPLITUDES**

The in-plane $I_{ac}$ and out-of-plane $I_b$ magnetic intensities for $SCCO-8$, as extracted from XYZ-PA on D7 data using Eq. 8, are shown in Supplementary Figure S5. From these maps, one sees that both magnetic components, in-plane and out-of-plane of the ladder a-c plane, are sizeable. For results obtained on both instruments, Supplementary Table S1 gives a summary of the measured magnetic intensities (as extracted from XYZ-PA) at different reciprocal space positions in $SCCO-5$ and $SCCO-8$ and the corresponding $I_{ac}$ and $I_b$ intensities in absolute units. The data were also systematically corrected by the quartz flipping ratios following the procedure given above for both 4F1 & D7 (we remind that the procedure is also described in [S10] for D7) . Note that, on 4F1, the measured **Q**-dependencies were systematically measured for negative and positive H values, and symmetrized by averaging the values of the magnetic intensity. The results in Supplementary Table S1 show that both in-plane $I_{ac}$ and out-of-plane $I_b$ magnetic components are not zero, leading systematically to a magnetic moment which is not pointing along a high symmetry direction, but typically is making a tilt with the direction perpendicular to the $CuO_2$ planes as it is observed in all superconducting cuprates [S2, S3].

|  |  | $I(0,0,1)$ | $I(1,0,1)$ | | $I(3,0,1)$ | |
| --- | --- | --- | --- | --- | --- | --- |
|  |  | 4F1 | 4F1 | D7 | 4F1 | D7 |
| $SCCO-8$ | $I_{mag}$(mbarn) | $0 \pm 5$ | $6 \pm 4.0$ | $16 \pm 10$ | $27.9 \pm 4.4$ | $36 \pm 15$ |
|  | $I_b$ | $-1 \pm 2$ | $3.1 \pm 1.7$ | $10 \pm 1$ | $12.7 \pm 1.5$ | $21 \pm 13$ |
|  | $I_{ac}$ | $1 \pm 2$ | $2.9 \pm 1.8$ | $6 \pm 1$ | $15.2 \pm 1.5$ | $15 \pm 8$ |
| $SCCO-5$ | $I_{mag}$(mbarn) | $11.4 \pm 2.9$ | $7.1 \pm 2.9$ | - | $7.4 \pm 2.4$ | - |
|  | $I_b$ | $4 \pm 1.7$ | $1.8 \pm 1.7$ | - | $5.4 \pm 1.4$ | - |
|  | $I_{ac}$ | $7.4 \pm 1.7$ | $5.3 \pm 1.7$ | - | $1.9 \pm 1.4$ | - |

Supplementary Table S1. Summary of the measured magnetic intensities for $SCCO-5$ and $SCCO-8$ on $D7$ and $4F1$.

## SUPPLEMENTARY NOTE 7: ABSENCE OF CHARGE DENSITY WAVE-LIKE INSTABILITY

A charge density wave has been reported in the two-leg ladders system, $SCCO$-$x$, from conductivity measurements that established a continuous phase diagram of the doped ladders [S11]. However, a contradictory result has been reported as resonant X-ray diffraction concluded that CDW order dominates at two particular $x$=0 and $x$=11 [S12], and is supposed to melt for other $x$, including the present $x$ = 5 and $x$=8.

In the pure compound $SCCO$, both ladders (ld) and chains (ch) display charge density waves (CDW), as reported by RXD and Neutron Diffraction studies [S13, S14]. For $CDW_{ld}$ and $CDW_{ch}$, the incommensurate propagation wavevector $q_{CDW}$ is given in super-space notations by $(H, K, L, L')$ with:

- $L = L_{integer} \pm 0.2\ r.l.u$ within the ladders
- $L' = L'_{integer} \pm 0.2\ r.l.u$, within the chains

Both CDW are characterized by a similar onset [S11]. For $SCCO-5$ and $SCCO-8$ compositions, optical conductivity measurement [S15, S16] report an onset of charge ordering within the ladders $T_{CDW}$ at ∼90K for $SCCO-5$ and 10K for $SCCO-8$.

In a neutron diffraction experiment, the charge order is detected owing the lattice distortion it induces. To detect the hallmark of a CDW, one has to look in the NSF channel in a PND study. During our PND experiments, we actually did not detect the hallmark of any CDW instability.

### Search for charge density wave in the ladders

According to literature [S11], $T_{CDW}$ should be at ∼90 K for the $SCCO-5$ sample. We performed two L-scans below and above that temperature. Supplementary Figure S6.a shows two scans along $(3, 1, L, 0)$, measured on 4F1 in the $NSF_X$ channel at T=8K and 150K. The difference between the two sets of measurements (Supplementary Figure S6.b) exhibits a featureless flat L-dependence only, pointing out the absence of any extra signal at $L$=1.2. Thus, the signal associated with a $CDW_{ld}$ (if any) falls below the threshold of detection of our measurement.

### Search for charge density wave in the chains

Supplementary Figure S6.c shows the same measurement within the chain subsystem, where one expects CDW scattering at $L'$=0.8, corresponding to $L$=1.14 using the ladder lattice parameter. Additionally, we collect a K-scan across $(3, 1, 0, 0.8)$, corresponding to $(3, 1, 1.14)$ in ladder notations. The differential intensity between 8K and 150K (Supplementary Figure S6.d) does not reveal any signature of $CDW_{ch}$.

The absence of evidence for a $CDW$ instability could also originate from the hole redistribution between the chains and the ladders upon $Ca$-doping, which leads to a change of the chains and ladders nuclear space group symmetry. These space group symmetry changes could affect $q_{CDW}$.

In addition, it is worth noticing that $T_{CDW}$ decreases with increasing the Ca content, while the ordering temperature $T_{mag}$ of the new short ranged magnetic signal keeps growing. This may highlight an interesting competition between the $q=0$ magnetism and the $CDW$, which could affect the development of the magnetic correlations.

### SUPPLEMENTARY NOTE 8: MAGNETIC PATTERNS AND RELATED STRUCTURE FACTORS

To describe the observed magnetic intensities, one needs to calculate the momentum dependence of the magnetic cross-sections (Eq. 1) for given magnetic patterns. Eq. 1 contains essentially two terms depending on **Q**: the magnetic form factor, $f(Q)$, which depends on the nature of the magnetic moments and the magnetic structure factor $F(\mathbf{Q})$ which is the Fourier transform of the given magnetic pattern. We discuss in this section several magnetic models and calculate $F(\mathbf{Q})$ for each of them.

Supplementary Figure S7.a shows the ladder unit cell in the a-c plane. The building block is given by a $CuO_2$ square plaquette with a characteristic length of the square lattice, $a_s = c$. Hereafter, all positions in real space are given in units of $a_s$. Note that as $a \sim 3a_s$, the size of ladder unit cell is approximately the size of 3 square plaquettes, although the ladder unit cell contains 4 inequivalent Cu atoms. Indeed, as shown in Supplementary Figure S7.a, the $Cu_4O_6$ unit cell is made of a first $Cu_2O_3$ ladder with two Cu sites on a rung at coordinates of (0,0) and (1,0). The second ladder is obtained by a translation of these coordinates by (3/2,1/2). This gives 4 distinct Cu sites distributed on 2 ladders. Each Cu site is at the center of a $CuO_2$ plaquette, with O sites at ($\pm 1/2$,0) and (0, $\pm 1/2$) around the Cu site. Note that there are only 6 distinct oxygen sites, since the two $CuO_2$ plaquettes on the same rung share one oxygen along the rung. Using the $CuO_2$ plaquette as a building block, one can decorate it with various magnetic patterns (Supplementary Figure S7.b-f). The magnetic dipoles can be related to a spin moment on Cu sites, spin or orbital moments on O sites, or orbital moments produced by loop currents (LC) between Cu and O sites or O sites only.

### Antiferromagnetic Cu spins

We consider a ladder where Cu spins are coupled antiferromagnetically (Supplementary Figure S8.a). This model comprises antiferromagnetic interactions along the ladder legs and rungs (due to superexchange interaction across the 180° oxygen bridge between $Cu$ ions). The existence of an antiferromagnetic order at long range is questionable. Indeed such a spin arrangement within ladders generates a frustration of the interladder magnetic interaction, due to the 90° oxygen bridges between neighboring ladders (Supplementary Figure S8.a). In this limit, one would actually expect the individual ladders to be in a non-classical state of singlets on each rung, forming spin dimers [S17]. This corresponds to a non magnetic ground state, without any magnetic static fingerprint observable. Here, our purpose is to look for the orign of a short range magnetism located at bragg positions. We therefore simply ask the question whether or not the Cu spins on a single ladder may lead to magnetic signal observed in neutron diffraction. Thus is at least valid for the isolated ladders limit of $SCCO$-5. As a consequence, one considers a set of independent antiferromagnetic ladders to compute the squared magnetic structure factor present in Eq. 1:

$$|F(\mathbf{Q})|^2 = |4\,sin(\pi L)\,sin(\pi \frac{H}{3})|^2 \qquad (12)$$

The two terms describe the antiferromagnetic coupling between 2 Cu spins along the leg and along the rung, respectively. This model breaks the lattice translation symmetry and should give a net magnetic contribution at half integer values of H and L which we did not observe during our experiment in $SCCO - 8$ (Supplementary Figure S8.b). It further rules out any magnetic scattering for integer H or L values which is at odds with our experimental measurements, where the magnetic scattering was observed at **Q**-positions of the form (H,0,1) with integer H. One can therefore eliminate a conventional Cu spin antiferromagnetism as the origin of the observed magnetic scattering.

### Magnetic moments on oxygen sites

Next, we consider a magnetic nematic model [S3, S18–S20] where O sites within a $CuO_2$ plaquette carry magnetic moments (spin or orbital) pointing in opposite directions for oxygen atoms located either along **a** or **c** directions with respect to the Cu site (Supplementary Figure S9). Once the ladders are decorated with such a magnetic nematic patterns, one observes 3 spins on O sites coupled ferromagnetically along **a**. They are coupled antiferromagnetically with 3 other spins translated by (1/2,1/2). This gives a squared magnetic structure factor, as follows:

$$|F(\mathbf{Q})|^2 = |2\,sin(\frac{\pi}{2}(\frac{H}{3} + L))\,(1 + 2\,cos(2\pi\frac{H}{3}))|^2 \tag{13}$$

The last term accounts for the ferromagnetic lines with 3 spins and the first term gives the **Q**-space relationship between neighboring lines with opposite spin directions. Such a magnetic pattern gives an extinction at (3,0,1) (Supplementary Figure S9.b). The model with magnetic moments on O sites (Supplementary Figure S9) then fails to account for the observation of a magnetic signal at (3,0,1).

### Loop current phases

We now discuss different magnetic patterns based on three distinct loop current models, shown in Supplementary Figure S7.d-f, all preserving the lattice translation symmetry.

#### *Single LC pattern*

Each $CuO_2$ square plaquette can be decorated by a single loop current (LC) pattern as shown in Supplementary Figure S7.d-f. For the three LC models, there are 4 different possibilities of putting the LC pattern around a given Cu site by making $\frac{\pi}{2}$ rotations. One can then write down the structure factor of a single LC pattern $F(\mathbf{Q}) \equiv A_\phi(\mathbf{Q})$ with $\phi=\{0, \frac{\pi}{2}, \pi, \frac{3\pi}{2}\}$, where $\phi$ denotes the angle of rotation for each pattern. One can conveniently describe all the different situations by introducing a local toroidal moment for a given $CuO_2$ plaquette $i$ : $\mathbf{\Omega}_i = \sum_j \mathbf{m}_j \times \mathbf{r}_j$, with $\mathbf{m}_j$ a magnetic moment and $\mathbf{r}_j$ its position with respect to the Cu site at the center of the $CuO_2$ plaquette [S21–S23]. Taking each of the four $\phi$ values, $\mathbf{\Omega}_i$ is pointing along the diagonal separating the clockwise and anticlockwise LCs. We consider three distinct LC states:

• **CC − $\theta_\mathbf{I}$** (Supplementary Figure S7.d): Theoretical works on copper oxide ladders predicted the appearance of a $CC - \theta_I$ phase with LCs in hole-doped $SCCO$ [S4, S5, S24]. On each $CuO_2$ plaquette, there are 4 LCs which generate staggered orbital moment at positions: $(\pm x_0, \pm x_0)$ with respect to a Cu site. $x_0 \sim 0.146$ is the coordinate of the triangle center of mass, where the orbital moment is assumed to be. This state is twofold degenerate with: $A_{\phi \pm \frac{\pi}{2}}(\mathbf{Q}) = -A_\phi(\mathbf{Q})$. The magnetic structure factor is independent of $\phi$ as:

$$A_\phi(\mathbf{Q}) = \pm 4\,sin(2\pi x_0 \frac{H}{3})sin(2\pi x_0 L) \tag{14}$$

$|A_\phi(\mathbf{Q})|^2$ for **CC − $\theta_\mathbf{I}$** is shown on Supplementary Figure S10.a (green line).

• **CC − $\theta_\mathbf{II}$** (Supplementary Figure S7.e): This state [S6, S25] breaks the inversion and time-reversal symmetry. In each square plaquette, it is made of two LCs turning clockwise and anti-clockwise and aligned along one diagonal ($\epsilon = +1$) or the other ($\epsilon = -1$). This state is fourfold degenerate with : $A_{\phi \pm \frac{\pi}{2}}(\mathbf{Q}) = -A_\phi(\mathbf{Q})$ and $A_{\phi \pm \frac{\pi}{2}}(\mathbf{Q}) = A_\phi(\mathbf{Q})$, corresponding to both orientations or domains ($\epsilon = \pm 1$). The two LCs produce 2 staggered orbital moments located with respect to the Cu site again at positions: $\pm(x_0, \epsilon x_0)$. The magnetic structure factor reads:

$$A_\phi(\mathbf{Q}) = 2\,sin(2\pi x_0(\frac{H}{3} + \epsilon L)) \tag{15}$$

$|A_\phi(\mathbf{Q})|^2$ for the two different **CC − $\theta_\mathbf{II}$** orientations ($\epsilon = \pm 1$) are shown on Supplementary Figure S10.a (dashed blue lines). The full blue line represents an averaged of both domains with equal population.

- **CC − $\theta_{III}$** (Supplementary Figure S7.f): this state corresponds to the LC pattern proposed to describe an ancillary phase associated within a spin liquid (mother) state [S7, S8], that for convenience we labeled here $CC - \theta_{III}$. At variance with the $CC - \theta_{II}$ state, the two LCs are rotated by $\frac{\pi}{4}$ and circulate between O sites only (Supplementary Figure S7.f). The two LCs produce 2 staggered orbital moments located with respect to the Cu site at positions: $\pm(x'_0, 0)$ for the *Horizontal* pattern or $\pm(0, x'_0)$ for the *Vertical* pattern, with $x'_0 = 0.5 - 2.x_0 = 0.208$ ($x'_0$ is again the center of mass of the LC triangle). For a ladder, the physics along the rung (*Horizontal*) and along the leg (*Vertical*) can be different, so that the fourfold degeneracy of the $CC - \theta_{III}$ state reduces to twofold only, with $A_{\phi+\pi}(\mathbf{Q}) = -A_\phi(\mathbf{Q})$. For both configurations, the magnetic structure factor is given by :

$$Horizontal: \ A_\phi(\mathbf{Q}) = 2\,sin(2\pi x'_0 \tfrac{H}{3}) \tag{16}$$

$$Vertical: A_{\phi+\frac{\pi}{2}}(\mathbf{Q}) = 2\,sin(2\pi x'_0 L) \tag{17}$$

$|A_\phi(\mathbf{Q})|^2$ are shown on Supplementary Figure S10.a for both configurations (dashed red lines).

*LCs correlations*

The next step is to establish how to correlate these patterns over the different sites of the ladder unit cell (Supplementary Figure S7.a). As discussed before, the $Cu_2O_3$ ladder contains 2 $CuO_2$ plaquette (with a diamond shape) which share an oxygen on a rung. For SCCO, the ladder unit cell contains 2 $Cu_2O_3$ ladders, with a translation from one to the other given by (3/2,1/2). Therefore, two types of LC correlations should be considered: i) the intra-rung correlations with the site shifted by (1,0) ii) the inter-ladder correlations with the site translated by (3/2,1/2). Both types of correlations contribute to the structure factor.

- **Intra-rung correlations**: For the **intra-rung**, one considers two patterns on each of both Cu sites of a rung, $A_\phi(\mathbf{Q})$ and $A_{\phi'}(\mathbf{Q})$. Defining $\varphi = \phi - \phi'$ and depending on the correlations, the pattern shifted by (1,0) is either identical ($\varphi = 0$ and $A_{\phi'}(\mathbf{Q}) = A_\phi(\mathbf{Q})$) or opposite ($\varphi = \pi$ and $A_{\phi'}(\mathbf{Q}) = -A_\phi(\mathbf{Q})$). In general, the intra-rung structure factor, $B_{\phi\phi'}(\mathbf{Q})$, can be written as:

$$B_{\phi\phi'}(\mathbf{Q}) = [A_\phi(\mathbf{Q}) + A_{\phi'}(\mathbf{Q})]cos(\pi \frac{H}{3}) + i[A_\phi(\mathbf{Q}) - A_{\phi'}(\mathbf{Q})]sin(\pi \frac{H}{3}) \tag{18}$$

$|B_{\phi\phi'}(\mathbf{Q})|^2$ is shown on Supplementary Figure S10.b for both correlations. Interestingly, this term gives zero structure factor at (3,0,1) for the opposite patterns ($\varphi = \pi$) at variance with the experimental results. This implies that both patterns are identical ($\varphi = 0$) and the intra-rung structure factor can be always simplified as:

$$B_\phi(\mathbf{Q}) = 2A_\phi(\mathbf{Q})cos(\pi \frac{H}{3}) \tag{19}$$

- **inter-ladder correlations**: The case of the **inter-ladder** is obtained in the same way defining two coupled LCs, $B_\phi(\mathbf{Q})$ and $B_{\phi'}(\mathbf{Q})$, shifted by (3/2,1/2). Again, defining $\psi = \phi - \phi'$ and depending on the correlations, the pattern shifted by (3/2,1/2) is either identical ($\psi = 0$ and $B_{\phi'}(\mathbf{Q}) = B_\phi(\mathbf{Q})$) or opposite ($\psi = \pi$ and $B_{\phi'}(\mathbf{Q}) = - B_\phi(\mathbf{Q})$) as for the intra-rung correlations. However, it is also possible that the LC shifted by (3/2,1/2) is aligned along a different diagonal than the one of the first ladder, $\psi = \pm\frac{\pi}{2}$ and $B_{\phi\pm\frac{\pi}{2}}(\mathbf{Q}) \neq B_\phi(\mathbf{Q})$. The inter-ladder structure factor can then be written as $C_{\phi\phi'}(\mathbf{Q})$ as:

$$C_{\phi\phi'}(\mathbf{Q}) = [B_\phi(\mathbf{Q}) + B_{\phi'}(\mathbf{Q})]cos(\frac{\pi}{2}(H+L)) + i[B_\phi(\mathbf{Q}) - B_{\phi'}(\mathbf{Q})]sin(\frac{\pi}{2}(H+L)) \tag{20}$$

$|C_{\phi\phi'}(\mathbf{Q})|^2$ is shown on Supplementary Figure S10.c for both correlations $\psi = 0$ and $\psi = \pi$. For in-phase ladders, one obtains the selection rule: $H + L = 2n$. The out-of-phase case ($\psi = \pi$) is ruled out by the experiments as it gives zero structure factor at (3,0,1).

Finally, the LCs magnetic structure factor $F(\mathbf{Q})$, present in Eq. 1, corresponds to (i) $A_\phi(\mathbf{Q})$ for an independent single pattern, (ii) $B_\phi(\mathbf{Q})$ for an independent ladder and (iii) $C_{\phi\phi'}(\mathbf{Q})$ for coupled ladders. It is worth to recall that the case (ii) corresponds to the results of the sample SCCO-5 and (iii) to SCCO-8, respectively.

### $CC - \theta_I$ like phase of LCs

The $CC - \Theta_I$ intra-rung pattern can be taken in-phase ($\varphi = 0$) as requested but the inter-ladder patterns are necessary out-of-phase as shown in Supplementary Figure S11.a as they share a current link along the diagonal of the inter-ladder small square. This gives the following $|F(\mathbf{Q})|^2$:

$$|F(\mathbf{Q})|^2 = |16\, sin(\frac{\pi}{2}(H+L))\, cos(\frac{\pi}{3}H)\, [sin(2\pi x_0 \frac{H}{3}) sin(2\pi x_0 L)]|^2 \tag{21}$$

According to the previous section, the third term (in brackets) corresponds to the $CC - \Theta_I$ pattern, the second term accounts for the ordering in-phase within the ladder and the first term the out-of-phase coupling between ladders. Such a structure factor gives rise to magnetic scattering extinction rules that do not account for our experimental observations (Supplementary Figure S11.b). For instance, it prohibits scattering when H and L are both odd, at variance with our observed magnetic scatterings at (1,0,1) and (3,0,1). Even in the case of SCCO-5 (independent ladder), it does not correspond to the results as $|F(\mathbf{Q})|^2 = 0$ for (0,0,1) where the magnetic signal is observed.

### $CC - \theta_{II}$ like phases of LCs

• *Uncorrelated ladders: $SCCO - 5$*

We first discuss the case of the independent ladders using the $CC - \theta_{II}$ pattern (Supplementary Figure S12.a). Within the ladder unit cell, the first ladder is decorated with in-phase pattern, whereas no LCs occur for the second ladder. $|F(\mathbf{Q})|^2$ reduces to a product of the in-phase intra-rung term times the magnetic pattern of Eq. 15:

$$|F(\mathbf{Q})|^2 = \sum_{\epsilon = \pm 1} \frac{1}{2} |4\, cos(\pi \frac{H}{3})\, sin(2\pi x_0 (\frac{H}{3} + \epsilon L))|^2 \tag{22}$$

Here, we assumed four possible domains with equal population. This structure factor (shown in Supplementary Figure S12.d) reproduces the SRM along the (H,0,1) line (see Fig. 5.b of the manuscript), while the magnetic correlations along **a** (perpendicular to the ladders) are confined within a single ladder.

• *Correlated ladders: $SCCO - 8$*

The magnetic structure factor is now given by Eq. 20. There are three different ways to couple the first and second ladders within the SCCO unit cell, corresponding to different phase shift $\psi$. The ladders couple in-phase ($\psi=0$) (Supplementary Figure S12.b), out-of-phase ($\psi = \pi$) or exhibit a crisscrossed coupling ($\psi = \pm \frac{\pi}{2}$) (Supplementary Figure S12.c). For the last case where $|B_{\phi'}(\mathbf{Q})| \neq |B_\phi(\mathbf{Q})|$ in Eq. 20, two different situations are possible to orient the magnetic patterns shifted by (3/2,1/2) denoted ($\delta = +1$) and ($\delta = -1$). Using the toroidal moment formalism (see section above), one can define $\bar{\mathbf{\Omega}} = \sum_i \mathbf{\Omega}_i$ the effective toroidal moment for the full SCCO unit cell. For the in-phase ladders ($\psi=0$), $\bar{\mathbf{\Omega}}$ remains along the same diagonal, whereas this vector is null for out-of-phase ladders. Interestingly, for the two crisscrossed cases, $\bar{\mathbf{\Omega}}$ points either along a rung, i.e the direction **a** ($\delta = +1$) or along a leg, i.e along the direction **c** ($\delta = -1$). The related squared structure factors, $|F(\mathbf{Q})|^2 = |C_{\phi\phi'}(\mathbf{Q})|^2$, are:

$$\psi = 0 : |F(\mathbf{Q})|^2 = \sum_{\epsilon=\pm 1} \frac{1}{2} |8\, cos(\frac{\pi}{2}(H+L))\, cos(\pi \frac{H}{3})\, sin(2\pi x_0 (\frac{H}{3} + \epsilon L))|^2 \tag{23}$$

$$\psi = \pi : |F(\mathbf{Q})|^2 = \sum_{\epsilon=\pm 1} \frac{1}{2} |8\, sin(\frac{\pi}{2}(H+L))\, cos(\pi \frac{H}{3})\, sin(2\pi x_0 (\frac{H}{3} + \epsilon L))|^2 \tag{24}$$

$$\psi = \pm \frac{\pi}{2},\, \delta = \pm 1 : |F(\mathbf{Q})|^2 = \{|2sin(2\pi x_0(\frac{H}{3} + L)) + \delta\, 2sin(2\pi x_0(\frac{H}{3} - L))|^2\, cos^2(\frac{\pi}{2}(H+L))$$
$$+ |2sin(2\pi x_0(\frac{H}{3} + L)) - \delta\, 2sin(2\pi x_0(\frac{H}{3} - L))|^2\, sin^2(\frac{\pi}{2}(H+L))\}$$
$$\cdot |2cos(\pi \frac{H}{3})|^2 \tag{25}$$

Among all $|F(\mathbf{Q})|^2$ (Supplementary Figure S12.d), the squared structure factor for the in-phase case (Supplementary Figure S12.b) is the one which reproduces the main features of our experimental results, namely: i) the absence of scattering at (0,0,1), ii) a scattering at odd H and L, iii) a stronger scattering at H=3 than at H=1. The crisscrossed $CC - \theta_{II}$ with the effective toroidal moment along the ladder ($\psi=\pm\frac{\pi}{2}$, $\delta = -1$), shown Supplementary Figure S12.c, is also consistent for the observed SRM, even if the difference of intensities between H=3 and H=1 is less pronounced in that case. Such a crisscrossed structure indicate that the effective toroidal moment should be along the ladder. Note that in bilayer cuprates $YBa_2Cu_3O_{6+x}$, the effective toroidal moment is found parallel to the underlying $CuO$ chains [S23].

<div style="text-align:center">$CC - \theta_{III}$ **like phases of LCs**</div>

<div style="text-align:center">• *Uncorrelated ladders:* $SCCO - 5$</div>

Supplementary Figure S13.a,d show SCCO unit cell when the first ladder is decorated with the same $CC - \theta_{III}$ pattern on each Cu site of the first ladder, while the second remains non magnetic. $|F(\mathbf{Q})|^2$ reads:

$$|F(\mathbf{Q})|^2 = |2\,cos(\pi\frac{H}{3})|^2\,|A(\mathbf{Q})|^2 \quad (26)$$

With $A(\mathbf{Q}) = 2\,sin(2\pi x'_0 \frac{H}{3})$ for the *Horizontal* pattern or $A(\mathbf{Q}) = 2\,sin(2\pi x'_0 L)$ for the *Vertical* one. Along (H,0,1) direction, the *Vertical*-$CC - \theta_{III}$ pattern accounts for the SRM observed in sample SCCO-5, with a typical modulation given by the term $|2cos(\pi\frac{H}{3})|^2$. This at variance, with the *Horizontal*-$CC - \theta_{III}$ pattern for which the scattering intensity cancels at H=0 which cannot explain the measured data.

<div style="text-align:center">• *Correlated ladders:* $SCCO - 8$</div>

Supplementary Figure S13.b,e show the case where the 2 ladders are decorated with the same $CC - \theta_{III}$ pattern ($\psi = 0$). This introduces in the squared structure factor an extra term $|2\,cos(\frac{\pi}{2}(H+L))|^2$ with respect to the uncorrelated case:

$$|F(\mathbf{Q})|^2 = |4\,cos(\pi\frac{H}{3})cos(\frac{\pi}{2}(H+L))|^2\,|A(\mathbf{Q})|^2 \quad (27)$$

The main features of our experimental results are: i) the absence of scattering at (0,0,1), ii) a scattering at odd H and L, iii) a stronger scattering at H=3 than at H=1. All these features are reproduced by both the *Horizontal* and the *Vertical* $CC - \theta_{III}$ pattern.

### SUPPLEMENTARY NOTE 9: MAGNETIC FORM FACTOR

We have considered various magnetic pattern within the $CuO_2$ plaquette involving spin on the Cu site, or spin/orbital moment on the oxygen sites and orbital moments originating from LCs. Another factor present in Eq. 1 is the magnetic form factor. Two different form factors can be considered here, either the isotropic magnetic Cu-form factor or the oxygen-one (which can be estimated from ref. [S26]). For LC states, electron are delocalized between several Cu and O sites or O sites only, but the exact form factor associated with the induced orbital moment remains unknown. Previous PND measurements in 2D cuprates [S27] suggested that both magnetic Cu- and O-form factor could be used. Supplementary Figure S14.a shows the **Q**-dependencies of $f(Q)^2$ for copper and oxygen. The **Q**-range of interest for our study is indicated by a shaded area, where $f_O^2(Q)$ varies of 47 % against 21% for $f_{Cu}^2(Q)$. In principle, since electrons are likely to be more delocalized for LCs, the magnetic O-form factor could be best suited to describe a fast decay of LCs magnetic signal as compared to the magnetic Cu-form factor. However, fitting the **Q**-dependence of the magnetic signal with either form factors gives good agreement with our data and the extracted magnetic moment amplitudes are, although different, of the same order of magnitude when considering the $O$ or $Cu$ form factors, as will be shown in the next section, Supplementary Table S2.

**SUPPLEMENTARY NOTE 10: MAGNETIC MOMENT AMPLITUDES AS EXTRACTED FROM DATA MODELING**

- *Out-of-plane magnetic moment: $m_b$*

As discussed above, among the different magnetic patterns of Supplementary Figure S7.b-f, some of the LCs-like phases can describe our experimental data. In principle, being confined in the (a,c) plane, classical LCs produce an orbital moment, $m_b \equiv m_{LC}$, pointing perpendicular to the LC plane. That corresponds to the magnetic intensity, $I_b$, that we have extracted from XYZ-PA (Supplementary Table S1). Using $I_b \propto m_b^2$, one can deduce the out-of-plane magnetic moment, $m_b \equiv m_{LC}$. Supplementary Figure S14.d-e show $I_b$ calibrated in $mbarns$ fitted by different models. The evolution of the magnetic intensity along (H,0,1) is rather different for SCCO-5 (Supplementary Figure S14.d) and SCCO-8 (Supplementary Figure S14.e). No correlation develops between the ladders in the former case and the magnetic scattering remains diffusive. For the latter case, magnetic correlations develop between the ladders. The scattered intensity $I_b$ can be described as:

$$SCCO-5: \qquad I_b(\mathbf{Q}) = r_o^2\, f(Q)^2\, m_b^2 |A(\mathbf{Q})|^2\, |2\cos(\pi\tfrac{H}{3})|^2 \qquad (28)$$

$$SCCO-8: I_b(\mathbf{Q}) = r_o^2\, f(Q)^2\, m_b^2 |A(\mathbf{Q})|^2\, |2\cos(\pi\tfrac{H}{3})|^2\, |2\cos(\tfrac{\pi}{2}(H+L))|^2 \qquad (29)$$

The Figs. S14.b-c show the effect of a **Q**-independent magnetic pattern, $|A(\mathbf{Q})|^2 = constant$, with in-phase coupling within a rung, on the one hand, and, between ladder, on the other hand. That reproduces the main features of the evolution of $I_b$ along (H,0,1). Next, in the Figs. S14.d-e, we add the evolution of the magnetic pattern of each LC phase, $|A(\mathbf{Q})|^2$. Supplementary Figure S14.d-e show the best fits of the PND data with $|A(\mathbf{Q})|^2$ corresponding to the different $CC - \theta_{II}$ and $CC - \theta_{III}$ LCs. The deduced values of $m_b$ are listed in supplementary Table S2.

- *In-plane magnetic moment: $m_{ac}$*

One can perform the same analysis for the full measured intensities $I_{mag}$ although, in principle, the classical LCs phase cannot account for the in-plane magnetic intensities reported in Supplementary Table S1. $I_{mag}$ is proportional to $m_\perp^2 = m_b^2 + m_{ac}^2$, with $m_{ac}^2 = (1 - |\tfrac{H}{Q}|^2)m_a^2 + |\tfrac{H}{Q}|^2 m_c^2$. As it combines magnetic moments along the ladder and along the rung, $m_{ac}$ varies with **Q** due to the orientation factors. For a H-scan along (H,0,1), the variations of these orientation factors, weighting $m_a^2$ and $m_c^2$ are given in Supplementary Figure S15. Unfortunately, the set of collected data for both samples remains insufficient to determine independently $m_a$ and $m_c$. As a consequence, one further constrains the fits by enforcing $m_a = m_c$. This simple assumption eliminates the Q-dependence of $m_{ac}$.

Supplementary Figure S15.b-c show the fit of the PND data for samples $SCCO-5$ and $SCCO-8$ using the same functions as in Supplementary Figure S14.d-e. One sees that the same LCs structure factor properly account for the full measured magnetic intenties. The resulting magnetic moments $m_\perp$ are listed in Supplementary Table S2. From these results, for both samples, one finds $\mathbf{m} = (m_a, m_b, m_a)$ with typically $|m| \simeq 0.09 \mu_B$: that value depends on the specific LCs pattern considered via $|A(\mathbf{Q})|^2$. That gives as well a tilt angle of 55° of the magnetic moment with respect to the **b** axis.

| $f(Q)^2$ | LC pattern | $m_b(x=8)$ | $m_b(x=5)$ | $m_\perp(x=8)$ | $m_\perp(x=5)$ |
|---|---|---|---|---|---|
| Cu | $CC - \theta_{II}$ | 0.05 | 0.05 | 0.07 | 0.07 |
| O | $CC - \theta_{II}$ | 0.06 | 0.05 | 0.08 | 0.07 |
| Cu | $CC - \theta_{III}$ Horizontal | 0.03 | 0.04 | 0.05 | 0.05 |
| O | $CC - \theta_{III}$ Horizontal | 0.05 | 0.06 | 0.08 | 0.08 |
| Cu | $CC - \theta_{III}$ Vertical | 0.03 | 0.04 | 0.05 | 0.06 |
| O | $CC - \theta_{III}$ Vertical | 0.05 | 0.05 | 0.08 | 0.08 |

Supplementary Table S2. Out-of-plane magnetic moment $m_b$ ($\mu_B$) and measured magnetic moment $m_\perp = \sqrt{m_b^2 + m_{ac}^2}$ ($\mu_B$) for both samples $SSCO-x$. The values are shown for both Cu and O form factors and for the different LCs magnetic patterns. Typical error on the estimation of the moment is 0.01 $\mu_B$.

# SUPPLEMENTARY NOTE 11: ORIGIN OF THE PLANAR COMPONENT

In the original model proposed by C.M. Varma, LCs are confined within the $CuO_2$ planes. This should generate only orbital magnetic moments perpendicular to the ladder planes, which is at variance with the experimental observation where an extra in-plane magnetic scattering is reported. Then, it was suggested that the ground state could not be solely made of one of the four orthogonal $CC-\theta_{II}$ states, but could rather emerge by their quantum superposition [S21, S22]. Within that framework, the degree of quantum admixture shows up in PND measurements in the form of an extra magnetic scattering that looks like that originating from an effective magnetic planar component. In $Cu_2O_3$ ladders, LCs settle in at lower temperature where quantum effects might be larger than thermal fluctuations. This makes the proposal of quantum effect at the origin of the planar magnetic scattering an interesting scenario.

Alternatively, it was proposed in superconducting cuprates that the planar component arises from LCs running over the $CuO_6$ octahedron [S2, S28–S30]. Indeed, the cuprates, where the q=0 magnetism was observed in monolayer and bilayer materials, are all containing $CuO_6$ octahedron with an apical oxygen site. In the $CuO_2$ layers, the Cu site is located at the center of either a $CuO_6$ octahedron or a $CuO_5$ pyramid. It was therefore proposed that LCs could delocalize on opposite edges of O-octahedra or O-pyramids, yielding a magnetic planar component. Whatever is the relevance of such a proposal for superconducting cuprates, it cannot hold for two-leg ladder cuprates, since there is no apical oxygen above the $CuO_2$ plaquette. For that kind of materials, LCs have to remain confined with the $Cu_2O_3$ ladders.

# SUPPLEMENTARY REFERENCES:



\end{document}